\documentclass{raa}            

\usepackage{graphicx,times}             
\usepackage{natbib}
\usepackage{amssymb,amsmath}
\bibpunct{(}{)}{;}{a}{}{,}

\usepackage[pagebackref=true]{hyperref}

\graphicspath{{./}{figures/}}

\begin{document}

\title{Close Major-merger Pairs at $z=0$:\\
Star-forming Galaxies with Pseudobulges}


   \volnopage{Vol.0 (20xx) No.0, 000--000}      
   \setcounter{page}{1}          

\author{
Chuan He (0000-0003-1761-5442, hechuan@nao.cas.cn)\inst{1,2}
\and
Cong Kevin Xu (0000-0002-1588-6700, congxu@nao.cas.cn)\inst{1}
\and
Ute Lisenfeld (0000-0002-9471-5423)\inst{3,4}
\and
Yu Sophia Dai (0000-0002-7928-416X)\inst{1}
\and
Taotao Fang (0000-0002-2853-3808)\inst{5}
\and
Jiasheng Huang (0000-0001-6511-8745)\inst{1}
\and
Wei Wang \inst{1,6}
\and
Qingzheng Yu (0000-0003-3230-3981)\inst{5}
}

\institute{Chinese Academy of Sciences South America Center for Astronomy, National Astronomical Observatories of China, Chinese Academy of Sciences, Beijing 100101, People's Republic of China
\and
School of Astronomy and Space Sciences, University of Chinese Academy of Sciences, Beijing 100049, People's Republic of China
\and
Departamento de Física Teórica y del Cosmos, Universidad de Granada, 18071 Granada, Spain
\and
Instituto Carlos I de Física Téorica y Computacional, Facultad de Ciencias, 18071 Granada, Spain
\and
Department of Astronomy, Xiamen University, Xiamen, Fujian 361005, People's Republic of China
\and
CAS Key Laboratory of Optical Astronomy, National Astronomical Observatories, Chinese Academy of Sciences, Beijing 100101, China\\
\vs\no
   {\small Received 08-Feb-2024; revised 20-Mar-2024; accepted 28-Mar-2024}
}

\abstract{
We present a study of star-forming galaxies (SFGs) with pseudobulges (bulges with S\'ersic index $\rm n < 2$) in a local close major-merger galaxy pair sample (H-KPAIR). With data from new aperture photometries in the optical and near-infrared bands (aperture size of 7\;kpc) and from the literature, we find that the mean Age of central stellar populations in Spirals with pseudobulges is consistent with that of disky galaxies and is nearly constant against the bulge-to-total ratio (B/T). Paired Spirals have a slightly lower fraction of pure disk galaxies ($\rm B/T \leq 0.1$) than their counterparts in the control sample. Compared to SFGs with classical bulges, those with pseudobulges have a higher ($>2\;\sigma$) mean of specific star formation rate (sSFR) enhancement ($\rm sSFR_{enh} = 0.33\pm0.07$ vs $\rm sSFR_{enh} = 0.12\pm0.06$) and broader scatter (by $\sim 1$\;dex). The eight SFGs that have the highest $\rm sSFR_{enh}$ in the sample all have pseudobulges. A majority (69\%) of paired SFGs with strong enhancement (having sSFR more than 5 times the median of the control galaxies) have pseudobulges. The Spitzer data show that the pseudobulges in these galaxies are tightly linked to nuclear/circum-nuclear starbursts. Pseudobulge SFGs in S+S and in S+E pairs have significantly ($>3\;\sigma$) different sSFR enhancement, with the means of $\rm sSFR_{enh} = 0.45\pm0.08$ and $-0.04\pm0.11$, respectively. We find a decrease in the sSFR enhancements with the density of the environment for SFGs with pseudobulges. Since a high fraction (5/11) of pseudobulge SFGs in S+E pairs are in rich groups/clusters (local density $\rm N_{1Mpc} \geq 7$), the dense environment might be the cause for their low $\rm sSFR_{enh}$.
}


   \authorrunning{He et al.}            
   \titlerunning{Pseudobulge in Paired Galaxies}  

   \maketitle
\section{Introduction}           
\label{sect:intro}
Broad and intensive studies have shown that galaxy interaction can induce star formation enhancement \citep{1972ApJ...178..623T,1978ApJ...219...46L,1987AJ.....93.1011K,1991ApJ...374..407X,1996ARA&A..34..749S}. Both simulations and observations have indicated that the existence and intensity of such enhancement is affected by many factors of the pair such as separation distance \citep{2004MNRAS.352.1081A,2004MNRAS.355..874N}, mass ratio \citep{2008MNRAS.384..386C}, environment \citep{2010MNRAS.407.1514E}, interaction phase \citep{2012MNRAS.426..549S}, orbital parameters \citep{2013MNRAS.433L..59P}. Meanwhile, some factors of the paired galaxies such as stellar mass \citep{2011ApJ...739L..40R}, redshift \citep{2014ARA&A..52..415M} and morphology \citep{1998ARA&A..36..189K} do affect the star formation behavior themselves, too. On the other hand, many quenching mechanisms are reported such as mass \citep{2010ApJ...721..193P}, environment \citep{2012ApJ...757....4P}, AGN feedback \citep{2013ARA&A..51..511K}, and bulge-to-total ratios \citep[B/T,][]{2009ApJ...707..250M,2014MNRAS.441..599B}. The combining effect of many factors especially when studying a big sample with a large range of parameters may bury potential results or even cause statistical bias. This inspires us to delve into the fundamental mechanism of the interaction-induced star formation. For instance, by investigating the kinematic asymmetry utilizing the recent IFS (integral field spectrograph) technique, \citet{2020ApJ...892L..20F} reveal that the strength of the ongoing tidal effect during certain merging phases is a more basic indicator to understanding the intermittent star formation enhancement in galaxy pairs than projected separations. Also, the SFR level is influenced by both the quantity of gas and its efficiency in forming stars. \citet{2019A&A...627A.107L} reveals that mergers in an earlier stage (i.e. pair) can take a lead in transferring atom gas to molecular gas, which makes the paired sample has a higher SFE when calculating the total gas, however, no significant SFE enhancement is found in their pair sample when calculating the molecular gas. On the other hand, the later stage mergers, who always show their strong strength on star formation activities by appearing as a ULIRG, can pool a large amount of molecular gas \citep[e.g.,][]{1991ApJ...366L...1S,1991ApJ...366L...5S}. By controlling the total gas content, \citet{2023ApJ...953...91L} found no significant enhancement on SFE in paired galaxies compared to the single galaxies. To conclude, it is still controversial whether and how galaxy interaction can induce star formation enhancement. To acquire solid evidence, one needs to select the sample and their controls carefully, and a detailed classification of their sample as well.

Far-infrared (FIR) observations by Spitzer \citep{2010ApJ...713..330X} and Herschel \citep{2016ApJS..222...16C} on a complete, unbiased close major-merger pairs sample selected from K$_s$-band \citet{2009ApJ...695.1559D} suggest that only SFGs in spiral-spiral (hereafter S+S) pairs have significantly enhanced specific star formation rate ($\rm sSFR = SFR/Mstar$), but not those in Spiral-Elliptical (hereafter S+E) pairs. This situation is also presented in \citet{2009ApJ...691.1828P,2010A&A...522A..33H,2010ApJ...713..330X,2019ApJ...882...14M}. \citet{2022ApJS..261...34H} studied the dependence of the interaction-induced specific star formation rate enhancement ($\rm sSFR_{enh} = \log (sSFR_{pg}) - \log (sSFR_{med,ctrl})$ where ``pg'' stands for paired galaxy and ``med,ctrl'' for the median of their control galaxies) on the bulge-to-total ratio (B/T). They found a negative dependence of the interaction-induced sSFR enhancement on the bulge-to-total ratio (B/T) and a significant ($>5\;\sigma$) enhancement only in paired SFGs with $\rm B/T < 0.3$. This is consistent with the results of theoretical simulations which predicted that a massive bulge can stabilize the disk and suppress the SFR during and after close encounters \citep{1996ApJ...464..641M, 2008MNRAS.384..386C, 2008A&A...492...31D}. However, it appears that SFGs with low B/T ratios, in particular the disky galaxies (B/T $\leq $ 0.1), have very diversified $\rm sSFR_{enh}$ with the value varying in a range of $\sim 2.5$ orders of magnitude, and even some of such SFGs have sSFR deficit.

In their analysis, \citet{2022ApJS..261...34H} separated pseudobulges (with S\'ersic index $\rm n < 2$) from classical bulges (with S\'ersic index $\rm n \geq 2$) and assigned $\rm B/T=0$ to galaxies with pseudobulges. This is because many pseudobulges found in two-component (bulge and disk) models \citep[such as those in][]{2022ApJS..261...34H} are misidentified bars, nuclear disks, and nuclear rings \citep{2004ARA&A..42..603K} which themselves may be triggered by interaction \citep{2019MNRAS.484.5192C, 2021MNRAS.502.2446E}. Meanwhile, the pseudobulges unrelated to nuclear star formation are mostly found in late-type spirals with relatively low B/T ratios \citep{2016ApJS..225....6K}, and assigning them to disky galaxies shall not introduce strong bias. However, bars and nuclear rings often have intermittent star formation activity and, when being caught in the ``off'' phase, may appear to be SFR-quenched \citep{2020MNRAS.499.1116F}. Also, some massive early-type spirals (such as S0 and Sa galaxies) with low gas content and low SFR may have large bars \citep{2017A&A...599A..43H}. 

Can the broader scatter of the $\rm sSFR_{enh}$ among disky SFGs be due to the diversity of galaxies with pseudobulges? The major scientific goal of this paper is to answer the above question and to investigate the roles played by pseudobulge SFGs regarding sSFR enhancement in paired galaxies. Different from \citet{2022ApJS..261...34H}, for galaxies with pseudobulges, we use the original B/T values instead of assigning $\rm B/T=0$. The SFG sample we use in this paper is introduced in Sect.~\ref{sect:sample}. In Sect.~\ref{sect:color} we present a study of the optical color-color diagrams measured in the inner part of the paired galaxies, particularly for galaxies with pseudobulges to separate those with bars, nuclear disks, and nuclear rings from normal pseudobulges of $n < 2$. In Sect.~\ref{sect:result} we present our science analyses and results. A discussion is carried out in Sect.~\ref{sect:discussion}. We conclude our paper in Sect.~\ref{sect:conclusion}. Throughout this paper, we adopt the $\Lambda$-cosmology with $\rm \Omega_m=0.3$ and $\Omega_\Lambda=0.7$, and $\rm H_0=70\;km\;s^{-1}\;Mpc^{-1}$.

\section{The sample} \label{sect:sample}
The sample we use in this paper is the same as in \citet{2022ApJS..261...34H}, which is the local close major-merger sample H-KPAIR \citep{2016ApJS..222...16C}. H-KPAIR is a subsample of the KPAIR which is a complete and unbiased K$_s$-band (Two Micron All Sky Survey, 2MASS) selected pair sample \citep{2009ApJ...695.1559D}. The H-KPAIR includes 44 Spiral+Spiral (S+S) pairs and 44 Spiral+Elliptical (S+E) pairs, all of which have spectroscopic redshifts in the range of $0.0067<z<0.1$, the projected separations in the range of $\rm 5\;h^{-1}\;kpc \leq s(p) \leq 20\;h^{-1}\;kpc$, the radial relative velocity $\rm \delta(Vz) < 500\;km\;s^{-1}$ and the K$_s$-band magnitude differences within 1 mag (corresponding to a mass ratio no greater than 2.5). The strict selection criteria of H-KPAIR gather the most enhanced closed major-merger pairs and makes it an ideal sample to explore the interaction effect to the paired galaxies. All galaxies in H-KPAIR have Herschel imaging observations in the six bands at 70, 110, 160, 250, 350 and 500, and their star formation rates (SFRs) are derived from the Herschel data \citep{2016ApJS..222...16C}. Seventy pairs have GBT 21cm HI observations \citep{2018ApJS..237....2Z}, and 78 spiral galaxies are observed by the IRAM 30m telescope for the CO emissions \citep{2019A&A...627A.107L}. The B/T ratios of H-KPAIR galaxies are taken from \citet{2022ApJS..261...34H} which are based on 2-D decompositions carried out using the SDSS $r$-band image. The stellar masses are also taken from \citet{2022ApJS..261...34H} who updated the formalism of \citep{2016ApJS..222...16C} by including a (g-r) color correction when estimating the mass from the K$_s$-band luminosity. Throughout this paper, we mainly focus on two subsamples of H-KPAIR: (1) all spiral galaxies Sect.~\ref{sect:color} and~\ref{subsec:pseudo-classical})
and (2) all star-forming galaxies (SFGs; Sect.~\ref{subsec:strong-enh} and~\ref{subsec:ss-se}). Following \citet{2016ApJS..222...16C} and \citet{2022ApJS..261...34H}, SFGs are defined by having $\rm log(sSFR/yr^{-1}) \geq -11.3$. There are 98 SFGs out of 132 Spirals (Table~\ref{tab:1}) in the H-KPAIR sample.

\section{Galaxy central colors} \label{sect:color}
We carry out aperture photometry for all galaxies in the H-KPAIR sample using a consistent round aperture in images of the SDSS $u$, $r$, $i$ and 2MASS K$_s$-bands. The $u$, $r$, $i$ images are generated using SDSS SAS mosaic tool\footnote{\url{https://dr12.sdss.org/mosaics/}}, which can stitch together several sky-subtracted, calibrated frames\footnote{\url{https://data.sdss.org/datamodel/files/BOSS_PHOTOOBJ/frames/RERUN/RUN/CAMCOL/frame.html}} to form a coherent image over a specified patch of sky using the {\sc sw}arp \citep{2002ASPC..281..228B}. For the K$_s$-band, the 2MASS images are resampled into the same pixel scale as the SDSS images of 0.396\arcsec/pix, and the background are subtracted with {\sc sw}arp. The median Full-Width-at-Half-Maximum (FWHM) of the point spread function (PSF) in the four bands are 1.53\arcsec, 1.32\arcsec, 1.26\arcsec and 2.9\arcsec, respectively.

For each galaxy, we use the same aperture in all four bands which corresponds to a projected diameter of 7\;kpc and is centered at its SDSS coordinates. Such a setup is to avoid overlap between apertures of the two member galaxies in a pair, given the minimal pair separation of $\rm 5\;h^{-1}\;kpc$ (corresponding approximately to $\rm 7\;kpc$ for $\rm H_0 = 70\; km\; s^{-1} \; Mpc^{-1}$). Also, the aperture diameters corresponding to $\rm 7\;kpc$ on our sources range from 4.2\arcsec to 53\arcsec, and most galaxies (141/176) have diameters larger than 2 times 2.9\arcsec, which is the worst resolution of the data (i.e. the $\rm FWHM = 2.9\arcsec$ for the K$_s$-band). Therefore no aperture correction is needed for any photometry.

The photometric errors of SDSS bands are calculated by combining the Poissonian error of the source flux, the RMS of background, and the calibration error\footnote{\url{https://data.sdss.org/datamodel/files/BOSS_PHOTOOBJ/frames/RERUN/RUN/CAMCOL/frame.html}}\footnote{\url{https://www.sdss4.org/dr12/algorithms/fluxcal/}}. Given the good S/N of our galaxies in the optical bands, errors in these bands are dominated by the calibration error (1\% for the $r$ and $i$-bands and 2\% for the $u$-band). For the 2MASS K$_s$-band, in addition to the error sources described above, the coadd noise caused by the resampling and smoothing when generating the Atlas Images is also considered\footnote{\url{https://irsa.ipac.caltech.edu/data/2MASS/docs/releases/allsky/doc/sec6_8a.html}}.

\begin{figure}[htb!] 
\includegraphics[width=\hsize]{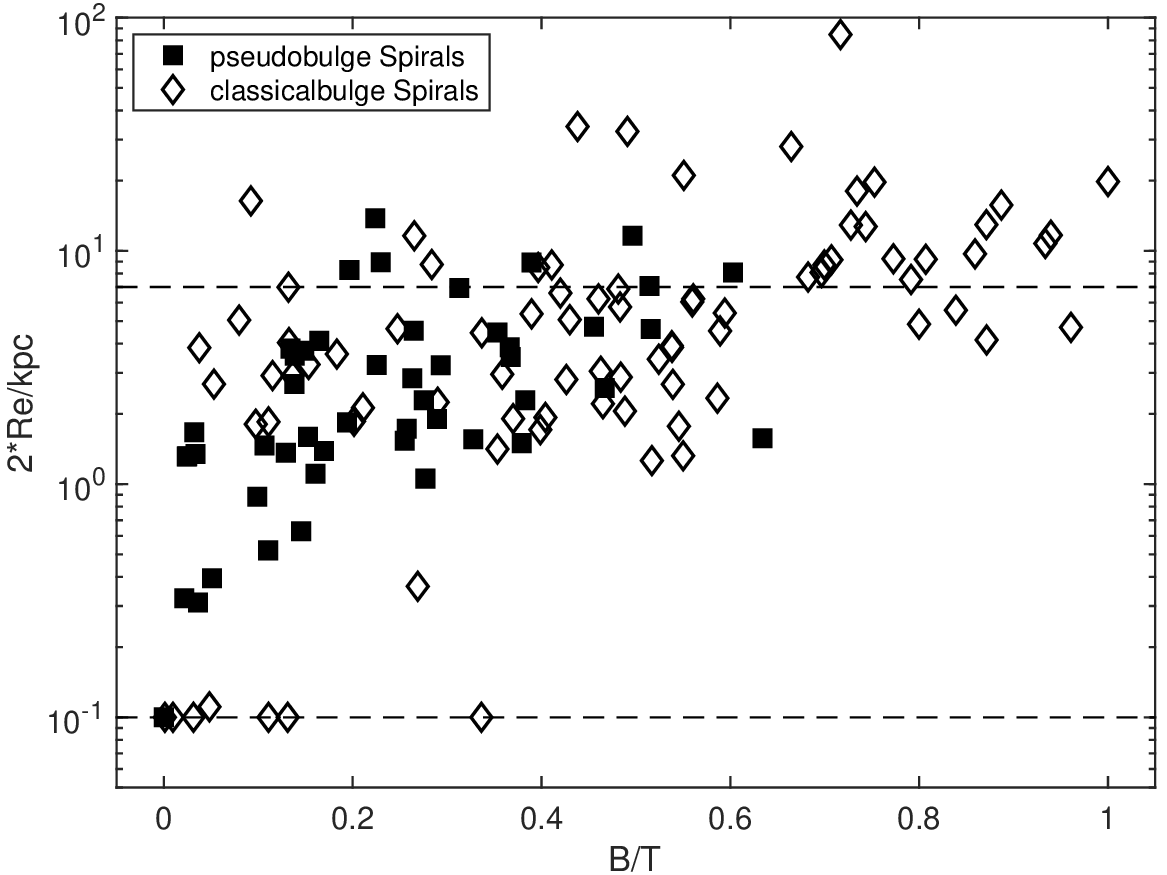}
\caption{Plot of bulge size ($\rm 2\times R_e$) versus B/T ratio of all spiral galaxies in H-KPAIR. The horizontal dashed line corresponds to the aperture size of our photometry ($\rm D=7\;kpc$) and a lower cut at $\rm 2 \times R_e=0.1\;kpc$ (sources with bulge sizes below this line are plotted on it). The filled squares represent galaxies with pseudobulges and the open diamonds galaxies with classical bulges.
\label{fig:Re-BT}}
\end{figure}

The spiral galaxies in the sample have a median bulge size of $\rm 2\times R_e = 3.6\;kpc$ ($\rm R_e$: effective radius of the bulge), and 26\% of them have $\rm 2\times R_e \geq 7\;kpc$. For the latter galaxies, the colors derived from the 7\;kpc aperture photometries provide information on the stellar population within the bulge. For the rest galaxies in the sample, the 7\;kpc aperture can include stellar radiations both in the bulge and a small portion of the inner disk, and their relative importance depends on the B/T ratio. As shown in Fig.~\ref{fig:Re-BT}, there is a significant correlation between bulge size and B/T (the Spearman's rank correlation coefficient $r_S = 0.55$ and the significance $p_S=5.6e^{-12}$).

To interpret the colors, predictions of broad-band fluxes are calculated using the Simple Stellar Population (SSP, or instantaneous-burst) model of \citet{2003MNRAS.344.1000B}, with the time after the instantaneous-burst $\tau$ (i.e. stellar age) and the attenuation ${\rm A}_v$ as the free parameters. The model uses the Padova~(1994) evolution tracks, the Chabrier~(2003) IMF with lower and upper mass cutoffs of 0.1–100 $\rm M_{\odot}$, and the solar metallicity ($\rm Z=0.02$). The extinctions are calculated following the formalism of \citet{1989ApJ...345..245C} for the standard diffuse interstellar medium. It is worth noting that the SSP model may be an over-simplification of the real stellar populations in question, and the model predictions on the $\tau$ and ${\rm A}_v$ have large uncertainties. Nevertheless, the model provides a framework to link the colors to the physical properties (e.g. the Age) of the stellar populations which allows us to investigate the difference between galaxies with pseudobulges and classical bulges using color-color diagrams. The results of our aperture photometry and the model fits are listed in Table~\ref {tab:2}.

\section{Results} \label{sect:result}
\subsection{Difference between pseudobulges and classical bulges} \label{subsec:pseudo-classical}

\begin{figure}[htb!]
\includegraphics[width=\hsize]{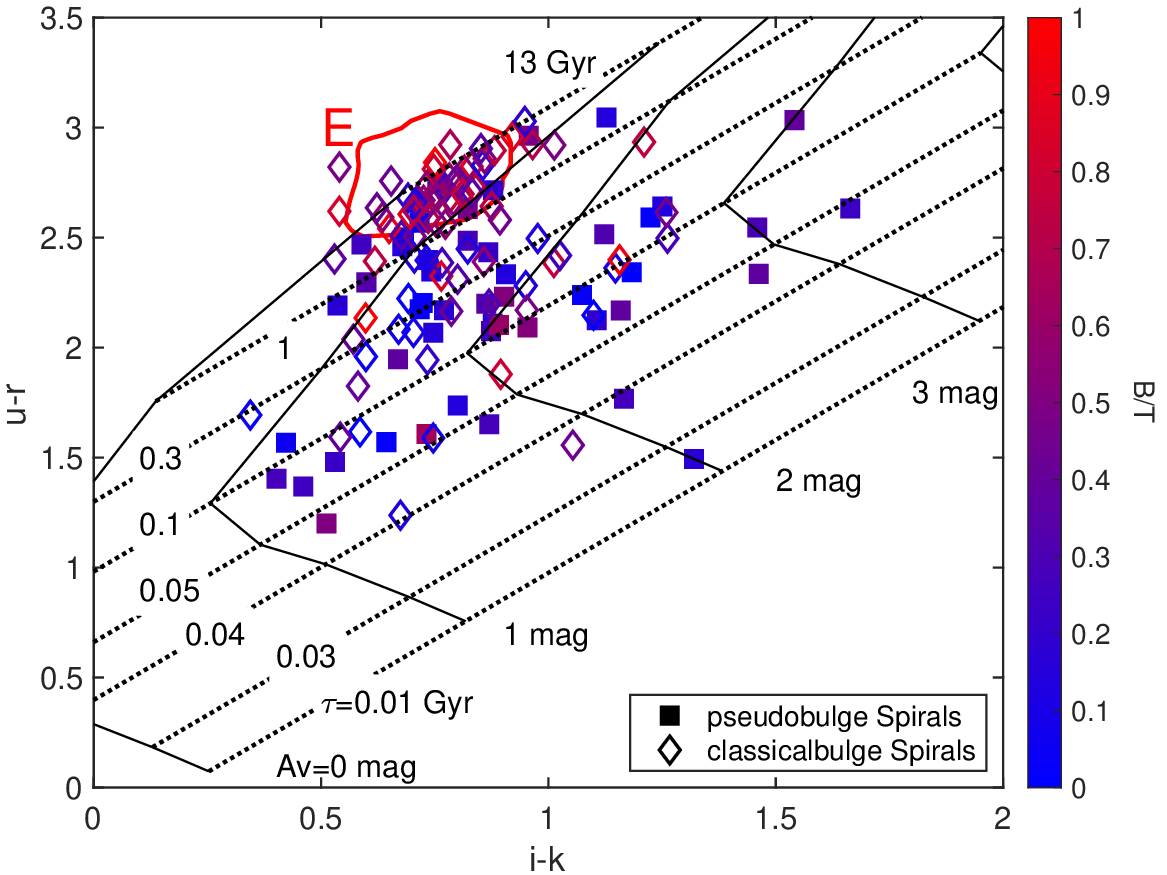}
\caption{The $u$-$r$-$i$-K$_s$ color-color diagram for Spirals with pseudobulges (filled squares) and classical bulges (open diamonds). The symbol color indicates the B/T ratio in \citet{2022ApJS..261...34H}. The red contour delineates the region of the distribution of H-KPAIR elliptical galaxies. The mesh grid outlines the time after the instantaneous-burst $\tau$ (i.e. stellar age) and attenuation ${\rm A}_v$ in the model. From bottom to top the dotted lines are $\tau$ equal to 0.01, 0.03, 0.04, 0.05, 0.1, 0.3, 1, 13\;Gyr, and from left to right the solid lines are ${\rm A}_v$ of 0, 1, 2, 3\; mag.
\label{fig:color-BT}}
\end{figure}

\begin{figure}[htb!]
\includegraphics[width=\hsize]{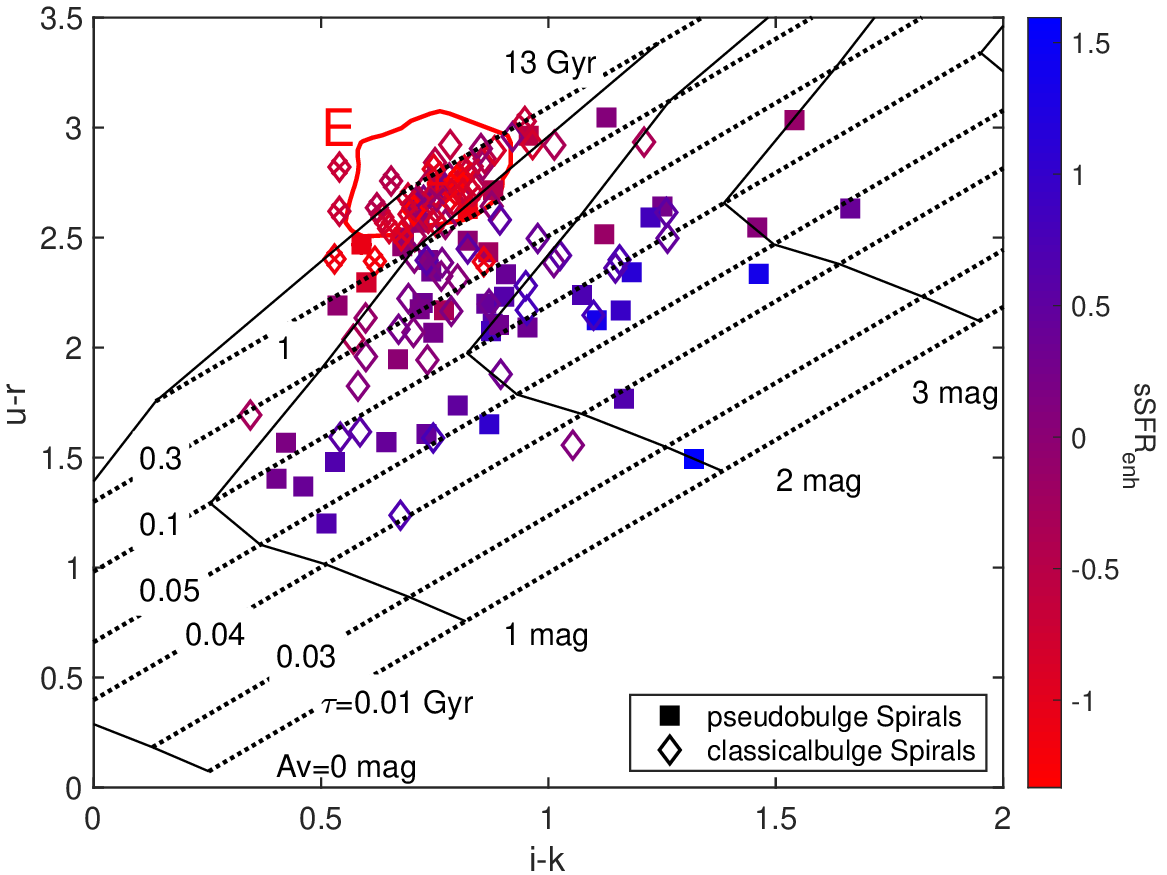}
\caption{Same $u$-$r$-$i$-K$_s$ color-color diagram as Figure~\ref{fig:color-BT}, but the color indicates the $\rm sSFR_{enh}$. The data points marked by crosses are SFR undetected in \citet{2016ApJS..222...16C}
\label{fig:color-enh}}
\end{figure}

Two $u$-$r$ versus $i$-K$_s$ color-color diagrams are presented in Fig.~\ref{fig:color-BT} and~\ref{fig:color-enh}, with the false color scale representing the B/T ratio and $\rm sSFR_{enh}$, respectively. The region outlined by the red contour (with a red label ``E'') in the upper-left corner of the diagram is where the elliptical galaxies (not plotted) are found. Most Spirals located in this region are those with relatively large classical bulges and low $\rm sSFR_{enh}$. The mesh grid in the diagram consists of lines for fixed values of $\tau$ and attenuation ${\rm A}_v$, respectively, as predicted by the SSP model (see Sect.~\ref{sect:color}). According to the model, galaxies in the region ``E'' have old stellar populations of $\rm Age > 1\;Gyr$ in their central part. For the remaining galaxies, most have their central stellar populations with Ages in the range of [0.1, 1] Gyr and very few with young Ages in the range of [0.01, 0.1] Gyr which are likely associated with nuclear/circum-nuclear starbursts. The range of ${\rm A}_v$ is between 0 and 3 mag, with most galaxies having $\rm 1\;mag < A_{\it v} < 2.5\;mag$.

\begin{figure}[htb!]
\includegraphics[width=\hsize]{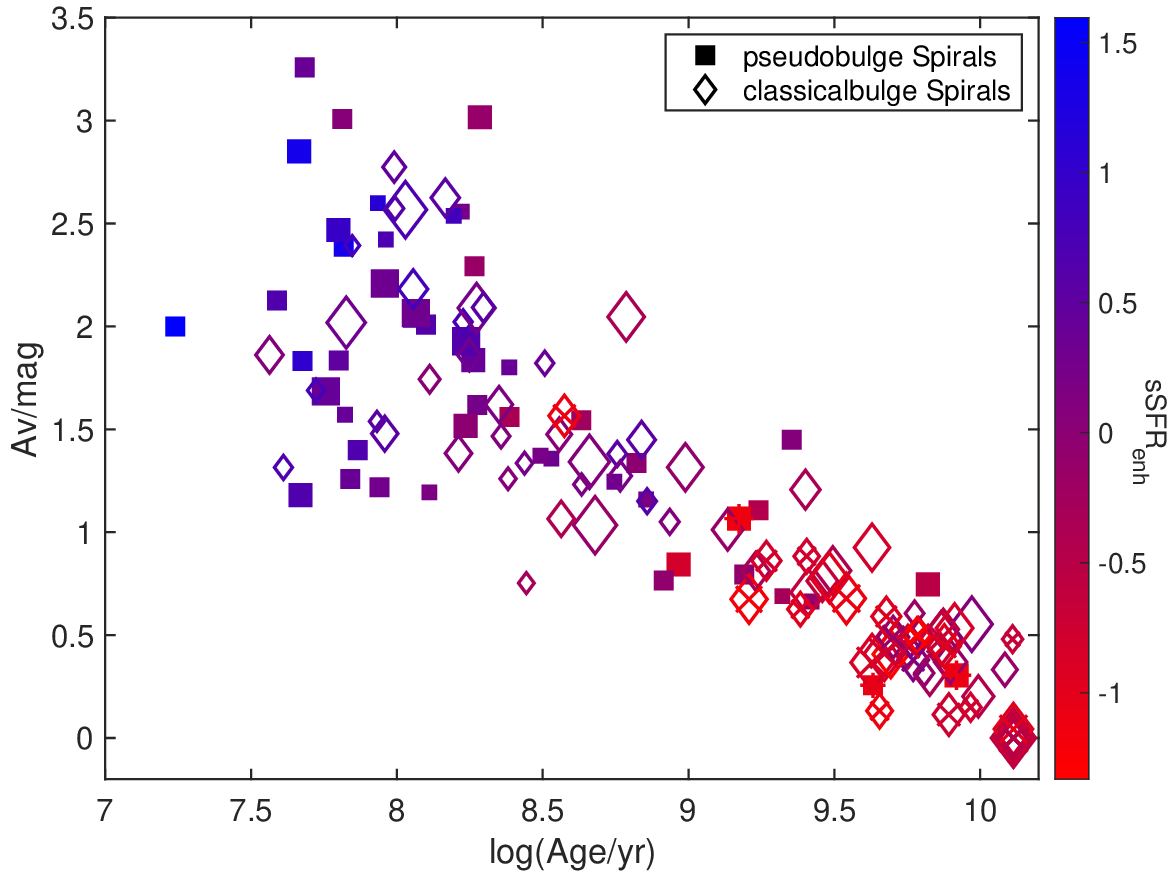}
\caption{Plot of the attenuation ${\rm A}_v$ vs.~age of the central stellar population for all spiral galaxies in the H-KPAIR. The color of symbols represents the $\rm sSFR_{enh}$, and the size of symbols represents the B/T ratio. The filled squares and the open diamonds represent pseudobulge galaxies and classical galaxies, respectively. The data points marked by crosses are SFR undetected in \citet{2016ApJS..222...16C}
\label{fig:age-Av}}
\end{figure}

We estimate the Age and ${\rm A}_v$ for each galaxy via interpolations among the model predictions (in the mesh grid) adjacent to its position in the color-color diagram. Galaxies with $\rm Age>13\;Gyr$ are regarded as $\rm Age=13\;Gyr$ and galaxies with ${\rm A}_v<0$ are regarded as ${\rm A}_v=0$. The plot of ${\rm A}_v$ versus Age (Fig.~\ref{fig:age-Av}) shows a strong anti-correlation between the two variables, which is understandable because younger stellar populations are usually associated with stronger star formation activity and higher dust attenuation, and older stellar populations with less star formation activity and dust attenuation. Indeed there are several galaxies with noticeable large classical bulge found to have high Av. We explain this by two aspects. Firstly, these galaxies have high $\rm sSFR_{enh}$, hence relatively high sSFR, though their control sample shall have low sSFR. The SFR of the sample is derived from Herschel observation \citep{2016ApJS..222...16C} and is mainly constrained by the dust temperature, then highly related to the attenuation. Secondly, our $u$-$r$-$i$-K$_s$ color which generated the Av is of the central part of the galaxy instead of the whole galaxy, and old galaxies have dust gathering in the central region (or we can say bulge) is reasonable.

To investigate whether interaction can affect the bulge performance, we select a new control sample (control-sample-2 hereafter). The selection of control-sample-2 is similar to the control sample in \citet[][, section 4]{2022ApJS..261...34H} except for without matching the B/T. The criteria of control-sample-2 go as follows:
\begin{enumerate}
\item Should be identified as a spiral in Galaxy Zoo \citep{2008MNRAS.389.1179L}.
\item Not in any interacting system, namely no neighbor galaxy in the SDSS database which has projected distance $\rm \leq 100\;kpc$ and observed redshift difference $\rm \leq 1000\;km\;s^{-1}$.
\item Has reliable B/T ratio, namely having $\chi^2/\nu < 2$ and no bad flag (flag bit 20 = 0) in \citet{2015MNRAS.446.3943M}.
\item The $\rm L_K$ matches that of the paired galaxy within 0.1~dex
\item Match of local density: we adopt a local density indicator $\rm N_{1Mpc}$, which is the count of galaxies brighter than $\rm M_{r} = -19.5$ and with redshifts differing less than $\rm 1000\;km\;s^{-1}$ from that of the target galaxy, in the surrounding sky area of radius = 1 Mpc (the count includes the target galaxy itself if it is brighter than $\rm M_{r} = -19.5$). By means of $\rm N_{1Mpc}$, we classify galaxies into 4 environmental categories: field ($\rm N_{1Mpc} \leq 3$), small group ($\rm 4 \leq N_{1Mpc} \leq 6$), large group ($\rm 7 \leq N_{1Mpc} \leq 10$), and cluster ($\rm N_{1Mpc}>10$). The control galaxy shall be in the same environmental category as the paired galaxy.
\item Has the closest redshift, among all qualified candidates, to that of the paired galaxy.
\end{enumerate}

\begin{figure}[htb!]
\includegraphics[width=\hsize]{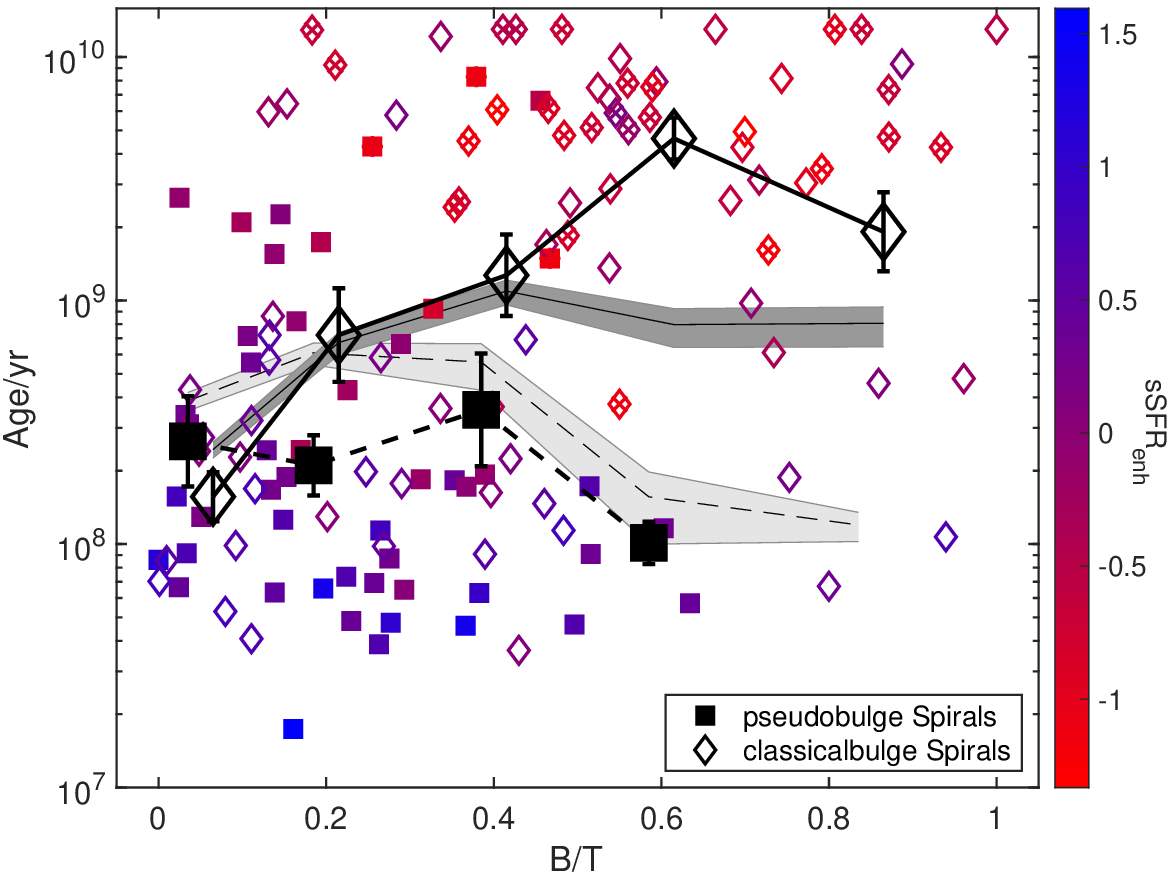}
\caption{Plot of Age vs.~B/T ratio for Spirals with pseudobulges (small filled squares) and with classical bulges (small open diamonds). The color of symbols represents $\rm sSFR_{enh}$. The large black squares and large black diamonds represent the means of pseudobulge galaxies and of classical-bulge galaxies in the five B/T ratio bins, respectively. The error bars show the 1-$\sigma$ errors of the means. The lines with light and dark shaded areas represent the means and their errors of the two kinds of galaxies in control-sample-2. The data points marked by crosses are SFR undetected in \citet{2016ApJS..222...16C}
\label{fig:age-BT}}
\end{figure}

In Fig.~\ref{fig:age-BT} we compare the Age and B/T ratio of the paired galaxies which are separated into two subsamples: (1) galaxies with pseudobulges and (2) those with classical bulges. Also plotted are the means of the Age for Spirals in H-KPAIR/control-sample-2 with error bars in the five B/T bins for the two subsamples: (1) $\rm B/T\leq0.1$, (2) $\rm 0.1 < B/T \leq 0.3$, (3) $\rm 0.3 < B/T \leq 0.5$, (4) $\rm 0.5 < B/T \leq 0.7$, (5) $\rm B/T>0.7$. It shows that, on average, the Age of the central stellar population of galaxies with classical bulges increases with the B/T ratio. This is indeed anticipated because the dominance of the bulge to central colors increases with the B/T ratio (Fig.~\ref{fig:Re-BT}), and in general stellar populations in classical bulges are older than those in disks \citep{2006MNRAS.368..414D,2016ApJS..225....6K}. In contrast, for galaxies with pseudobulges, the Age of central stellar populations is rather flat against the B/T ratio, namely the stellar populations have about the same Ages as those in the inner disks of the disky galaxies (with $\rm B/T \sim 0$). This confirms that pseudobulges are different from classical bulges, and the former are dominated by the ``disk phenomena'' such as bars and nuclear disks/rings as assumed in \citet{2022ApJS..261...34H}. However, some individual galaxies with pseudobulges have central stellar populations with old Ages of $\rm > 1\;Gyr$. Their pseudobulges are unlikely related to any central star formation activity, as indicated by the low star formation enhancement ($\rm sSFR_{enh} \lesssim 0$).  They have relatively low B/T ratios (most with $\rm B/T < 0.3$, all with $\rm B/T < 0.5$) and may be similar to those single late-type galaxies with relatively small pseudobulges found in \citet{2016ApJS..225....6K}. 

It is also interesting to note that a few galaxies with large classical bulges are located in the lower-right quadrant of Fig.~\ref{fig:age-BT}, namely their central stellar populations appear to have relatively young Ages. They may contain starbursts or post-starbursts in the central region which can outshine the old stellar populations and make the bulges ``blue''. An outstanding example of these galaxies is J13151726+4424255, which has a classical bulge with the $\rm B/T=0.94$, the $\rm Age=0.11\;Gyr$, and the $\rm sSFR_{enh}=0.70$. Compared to galaxies in the control sample, paired galaxies with large bulges, particularly those in the forth bin of the B/T ratio ($\rm 0.5 <  B/T \leq  0.7$), have higher Ages. Individual inspections of these galaxies indicate that most of them have elliptical companions (i.e. in SE pairs). Their star formation history might have been affected by the environment associated with the Ellipticals (see the discussion in Section 5).

\begin{figure}[htb!]
\includegraphics[width=\hsize]{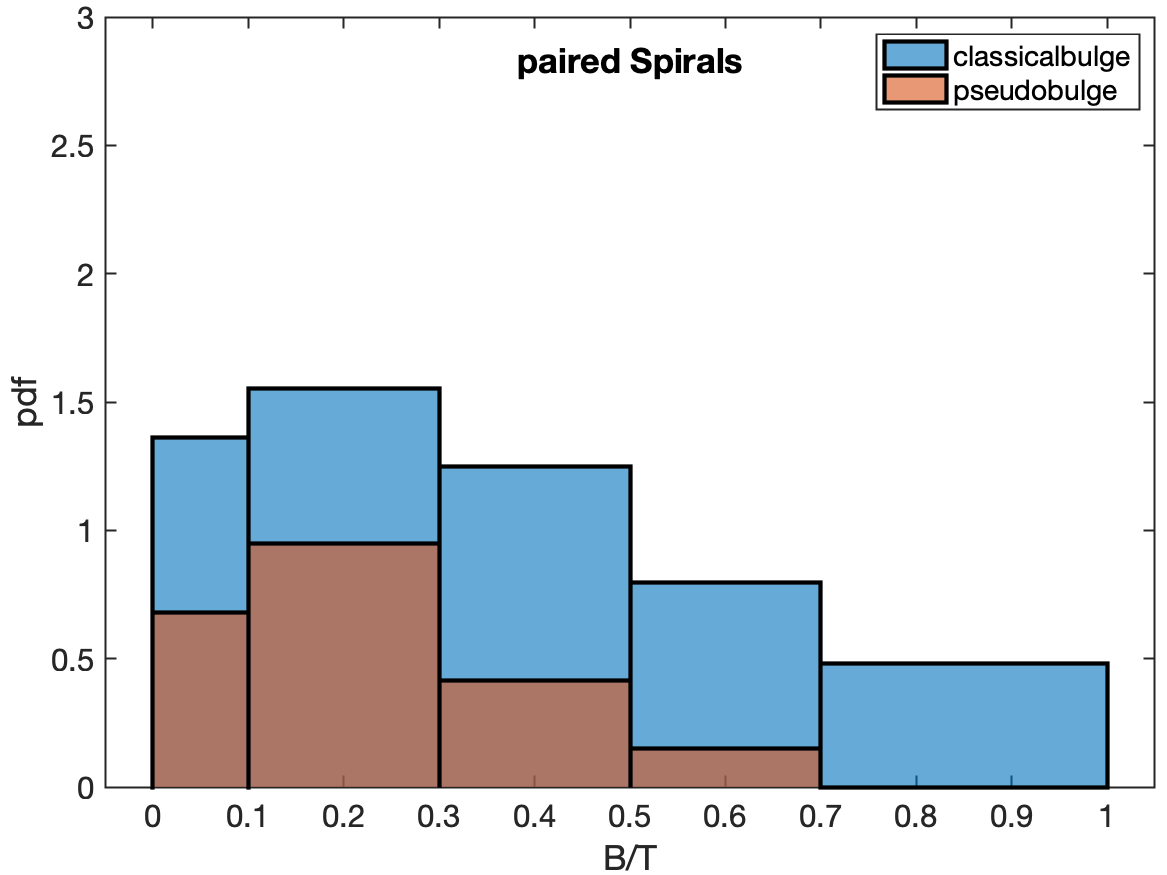}
\caption{Probability density function of pseudobulge Spirals and classical bulge Spirals in H-KPAIR.
\label{fig:hist-pair}}
\end{figure}

\begin{figure}[htb!]
\includegraphics[width=\hsize]{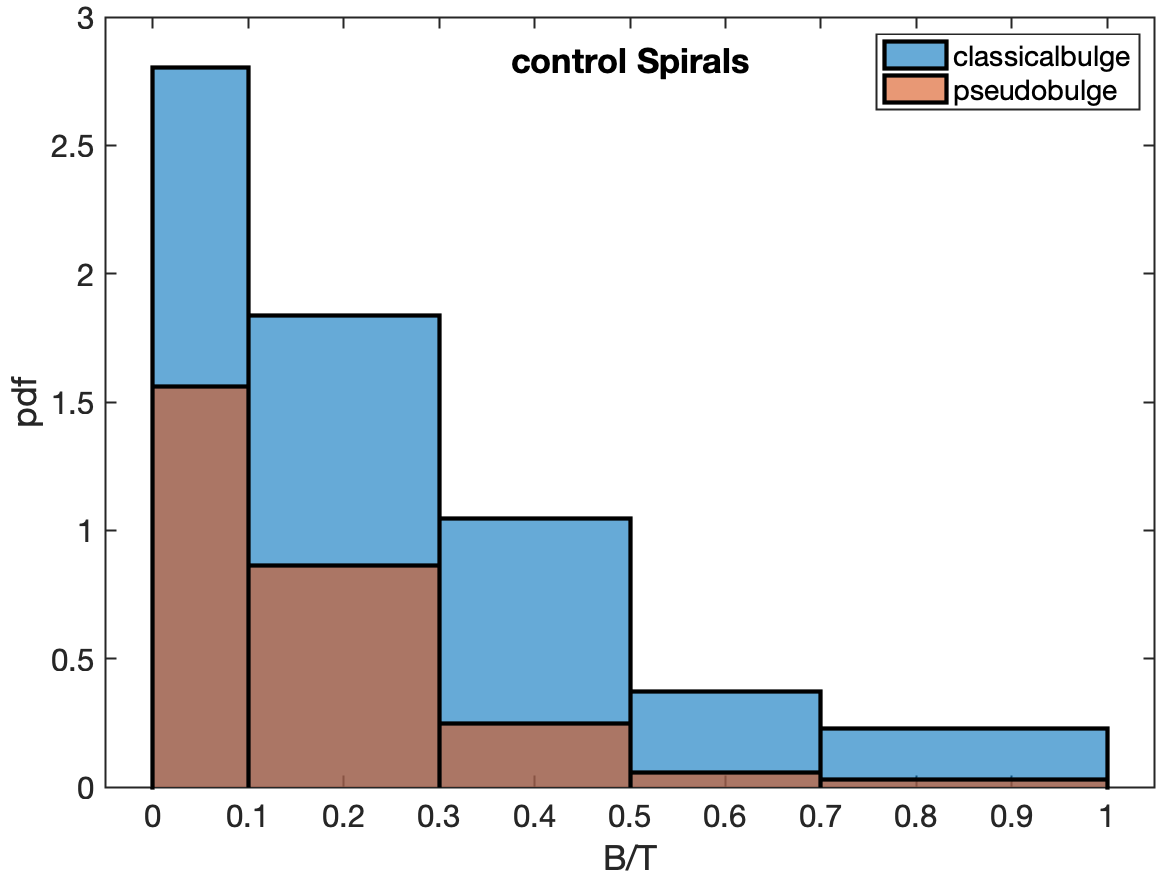}
\caption{Probability density function of pseudobulge Spirals and classical bulge Spirals in control-sample-2
\label{fig:hist-ctrl}}
\end{figure}

Fig.~\ref{fig:hist-pair}~and~\ref{fig:hist-ctrl} show the possibility density function (pdf) of classical bulges and pseudobulges in the five bins for SFGs in H-KPAIR and control-sample-2, respectively. The paired Spirals have lower fraction of pure disk galaxies (bin1, $\rm B/T\leq0.1$), we attribute this to some paired disky galaxies having nuclear/circum-nuclear starburst and their B/T ratio raising to higher B/T bins. Nevertheless, these galaxies have a flat profile and is identified as pseudobulge galaxies and are assigned $\rm B/T=0$ in \citet{2022ApJS..261...34H}

\subsection{Paired SFGs with strong enhancement} \label{subsec:strong-enh}

In Sect.~\ref{sect:sample} we defined SFGs by $\rm log(sSFR/yr^{-1}) \geq -11.3$ \citep[see also][]{2016ApJS..222...16C,2022ApJS..261...34H}. Most of the non-SFGs are SFR undetected (marked by crosses in Figure~\ref{fig:color-BT},~\ref{fig:age-Av} and~\ref{fig:age-BT}) in \citet{2016ApJS..222...16C}, and they are mostly early-type galaxies with large old classical bulge. Taking these galaxies into the $\rm sSFR_{enh}$ analysis shall not affect our results. From now on we focus on the SFGs subsample.

\begin{figure}[htb!]
\includegraphics[width=\hsize]{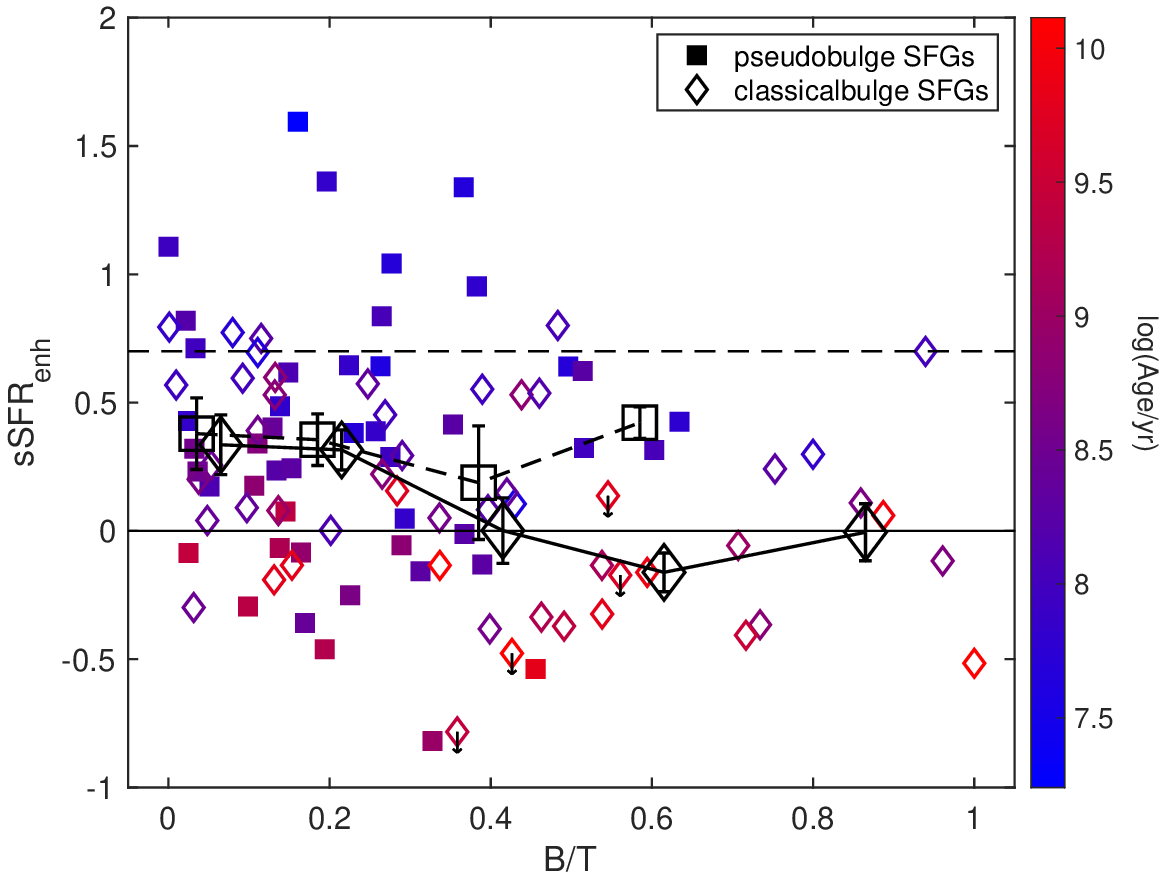}
\caption {Plot of $\rm sSFR_{enh}$ vs.~B/T ratio for SFGs with pseudobulges (small filled squares) and SFGs with classical bulges (small open diamonds). The color of symbols represents the Age of the central stellar population. The large black open squares and large black open diamonds represent the means of pseudobulge galaxies and of classical-bulge galaxies in the five B/T ratio bins, respectively. The error bars show the 1-$\sigma$ errors of the means. The horizontal solid line and dashed line mark $\rm sSFR_{enh}=0$ and $\rm sSFR_{enh}=0.7$, respectively.
\label{fig:enh-BT}}
\end{figure}

Fig.~\ref{fig:enh-BT} is a plot of $\rm sSFR_{enh}$ against B/T ratio, which is similar to Figure~9 of \citet{2022ApJS..261...34H} except for using the original B/T values (instead of $\rm B/T=0$) for galaxies with pseudobulges, and the 5 bins are divided the same as in Figure~\ref{fig:age-BT}. It shows a broad scatter of $\rm sSFR_{enh}$ for galaxies with pseudobulges, which spans about two orders of magnitude ($\sim 1$ orders magnitude larger than that for galaxies with classical bulges). This demonstrates that the broad scatter of $\rm sSFR_{enh}$ for disky galaxies (with $\rm B/T < 0.1$) found by \citet{2022ApJS..261...34H} is indeed due to galaxies with pseudobulges which are assigned $\rm B/T=0$ in that paper. The mean $\rm sSFR_{enh}$ of pseudobulge SFGs is rather constant against the B/T ratio, in contrast to that of SFGs with classical bulges which decreases with increasing B/T (Fig.~\ref{fig:enh-BT}; see also \citealt{2022ApJS..261...34H}). The mean $\rm sSFR_{enh}$ of SFGs with pseudobulges ($\rm sSFR_{enh} = 0.33\pm 0.07$) is higher than that of SFGs with classical bulges ($\rm sSFR_{enh} = 0.12\pm 0.06$) at $2\;\sigma$ level.

\begin{figure}[htb!]
\includegraphics[width=\hsize]{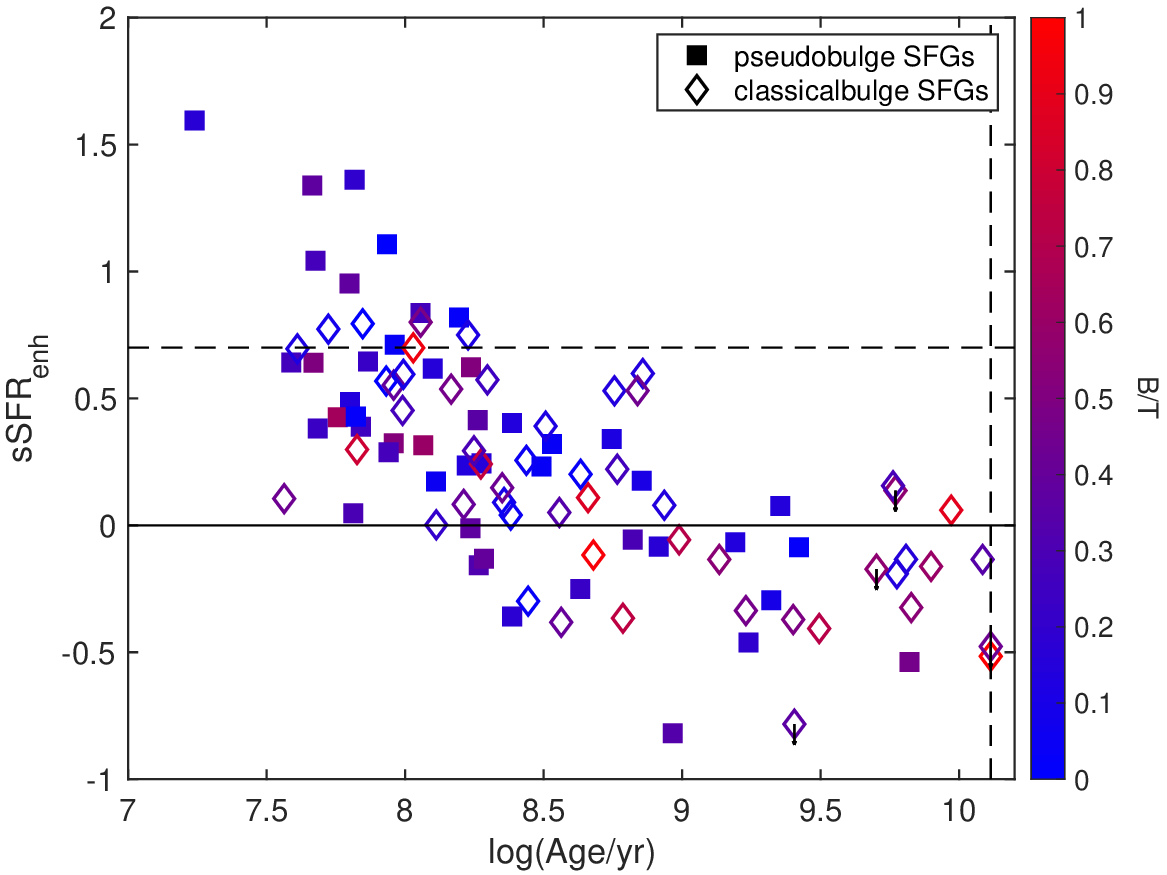}
\caption{Plot of $\rm sSFR_{enh}$ vs.~age of the central stellar population SFGs with pseudobulges (filled squares) and SFGs with classical bulges (open diamonds). The color of symbols represents the B/T ratio. The horizontal solid line and dashed line mark $\rm sSFR_{enh}=0$ and $\rm sSFR_{enh}=0.7$, respectively. 
\label{fig:enh-Age}}
\end{figure}

The eight SFGs with the highest $\rm sSFR_{enh}$ all have pseudobulges (Fig.~\ref{fig:enh-BT}). In order to study the galaxies with the strong sSFR enhancement in detail, we concentrate on those having $\rm sSFR_{enh}>0.7$, corresponding to more than 5 times sSFR enhancement compared with the median of the matching controls. With this criterion, there are 13 strongly enhanced SFGs in our sample, nine with pseudobulges and four with classical bulges (see Fig.~\ref{fig:enh-BT} and Table~\ref{tab:3}). Fig.~\ref{fig:enh-Age} shows that their central stellar populations all have Ages younger than 200\;Myrs, namely are dominated by stars formed recently. Excluding these galaxies, the rest of our sample have a mean $\rm sSFR_{enh} = 0.11\pm0.04$, which is only marginally significant at $2.8\;\sigma$ level. Statistically, SFGs with strong enhancement represent 13\% (=13/98) of SFGs in our sample and 10\% (=13/132) of all spiral galaxies in the H-KPAIR sample.

\begin{figure}[htb!]
\includegraphics[width=\hsize]{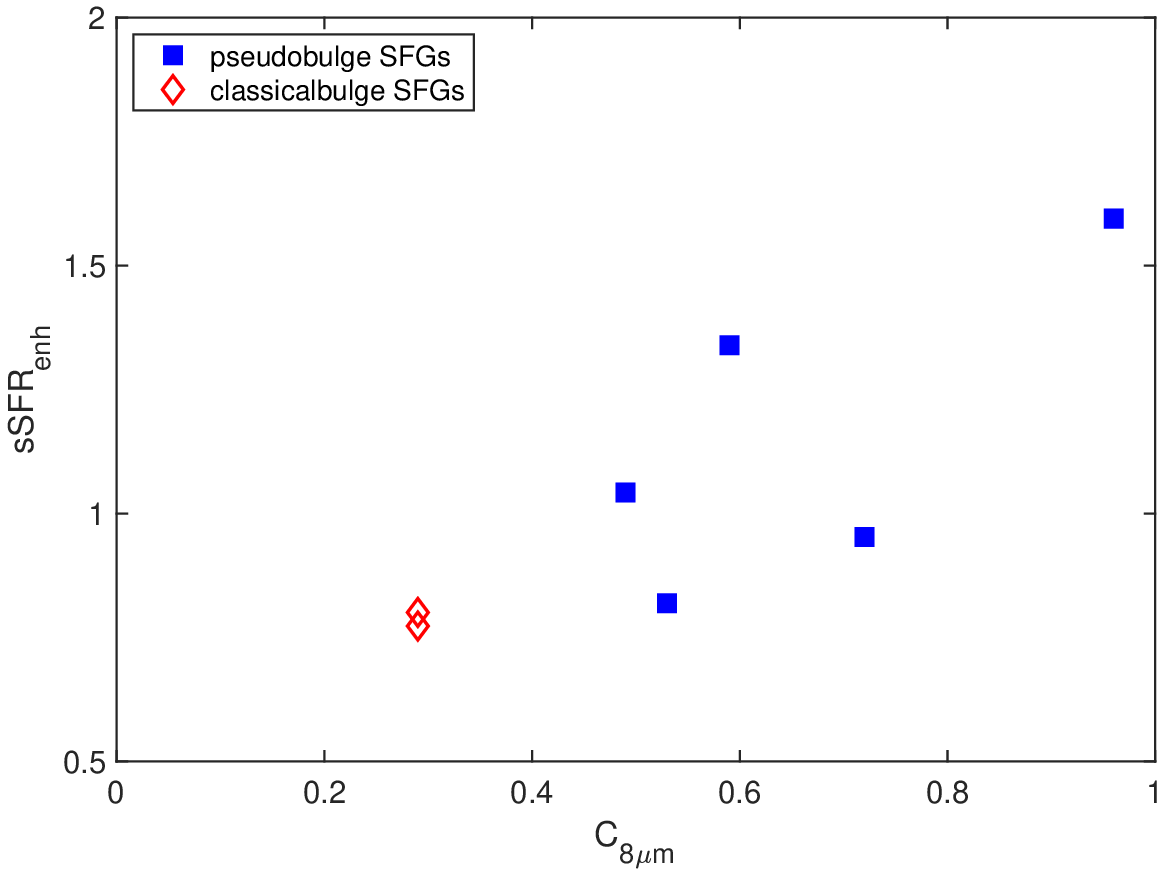}
\caption{Plot of $\rm sSFR_{enh}$ vs.~the nuclear-to-total ratio of the $\rm 8 \mu m$ emission $\rm C_{8\mu m}$. 
\label{fig:enh-c8}}
\end{figure}

Seven of the 13 galaxies with strong enhancement have Spitzer observations and consequently, the $\rm 8 \mu m$ flux densities of both the nuclear emission (aperture size: $\rm D=4\;kpc$) and the total emission \citep{2010ApJ...713..330X}. In Fig.~\ref{fig:enh-c8} we plot $\rm sSFR_{enh}$ against $\rm C_{8\mu m}$ for these galaxies, where $\rm C_{8\mu m} = f_{8\mu m, nuclear}/f_{8\mu m, total}$ is the nuclear-to-total ratio of the $\rm 8 \mu m$ emission. $\rm C_{8\mu m}$ is an indicator of the nuclear concentration of star formation in a galaxy. The plot shows a significant linear correlation between $\rm sSFR_{enh}$ and $\rm C_{8\mu m}$ (the Pearson's linear correlation coefficient $\rm r_P=0.81$  and the significance $\rm p_P=0.03$) for these galaxies. Among the seven SFGs in the plot, all the five with pseudobulges have relatively high nuclear concentration ($\rm C_{8\mu m} \geq 0.49$) whereas, for the two SFGs with classical bulges, the star formation occurs mostly outside the nuclear region ($ C_{8\mu m} < 0.30$). Albeit the small sample size, the plot indicates that the pseudobulges in galaxies with strong enhancement are tightly related to nuclear/circum-nuclear starbursts. Given that most galaxies with strong enhancement have pseudobulges, our result suggests that nuclear/circum-nuclear starburst may be the dominant mode for strong tidally induced star formation in paired galaxies. On the other hand, for a low fraction of paired SFGs (particularly those with classical bulges), the tidally induced star formation may be widely distributed over the entire disk. These results are consistent with the literature which shows that nuclear/circum-nuclear starbursts are common among paired galaxies with strong signs of interaction while wider spread star formation enhancement is also observed in some of them \citep{1985AJ.....90..708K,1987AJ.....93.1011K,2019ApJ...881..119P,2021ApJ...909..120S}.

\begin{figure*}[htb!]
\includegraphics[width=\hsize]{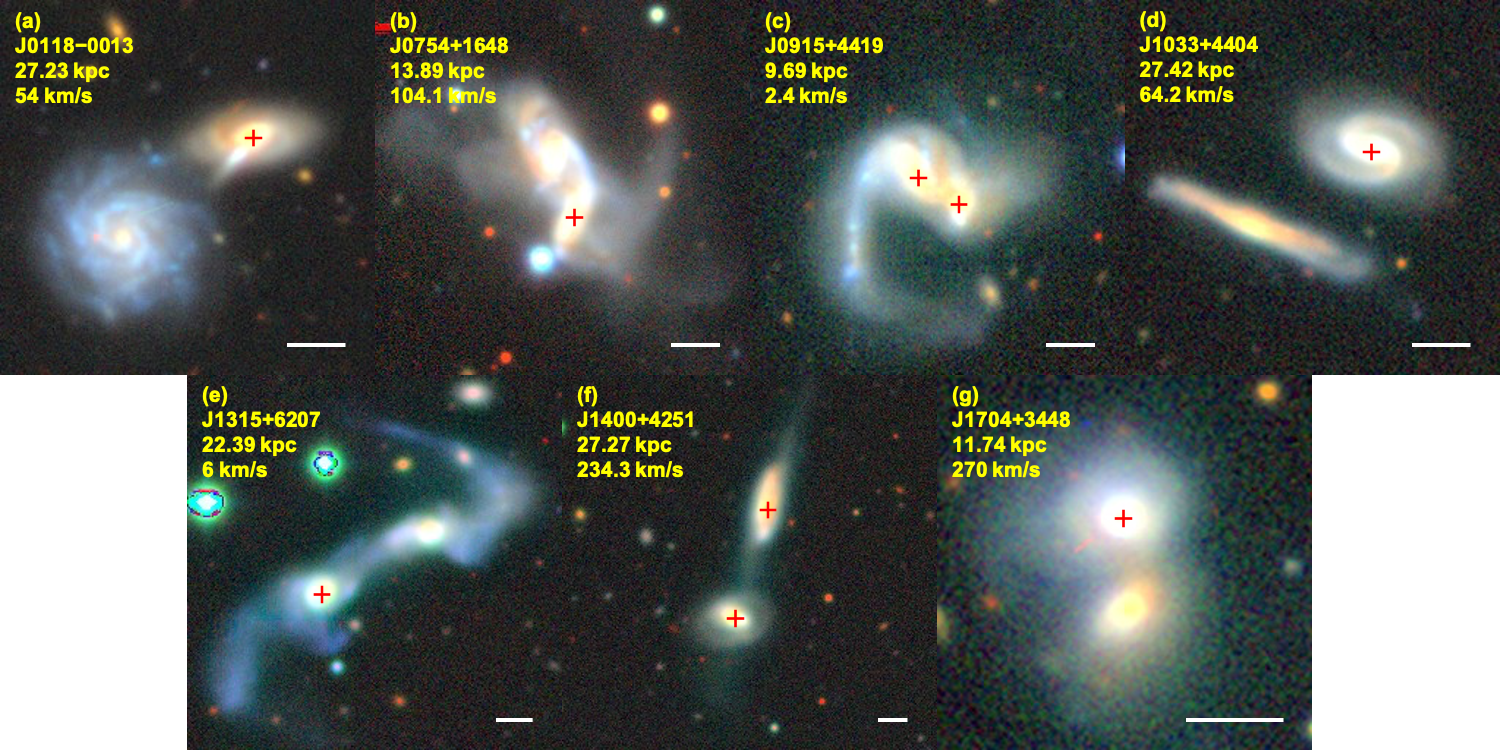}
\caption{Dark Energy Spectroscopic Instrument (DESI) images of nine pseudobulge galaxies with strong star formation enhancement. The strongly enhanced galaxies are marked by red plus signs. The white dash on the right-bottom corner of each stamp illustrates the angular scale of $10\arcsec$. The pair separation and velocity difference are provided as well.
\label{fig:pseudo}}
\end{figure*}

\begin{figure*}[htb!]
\includegraphics[width=\hsize]{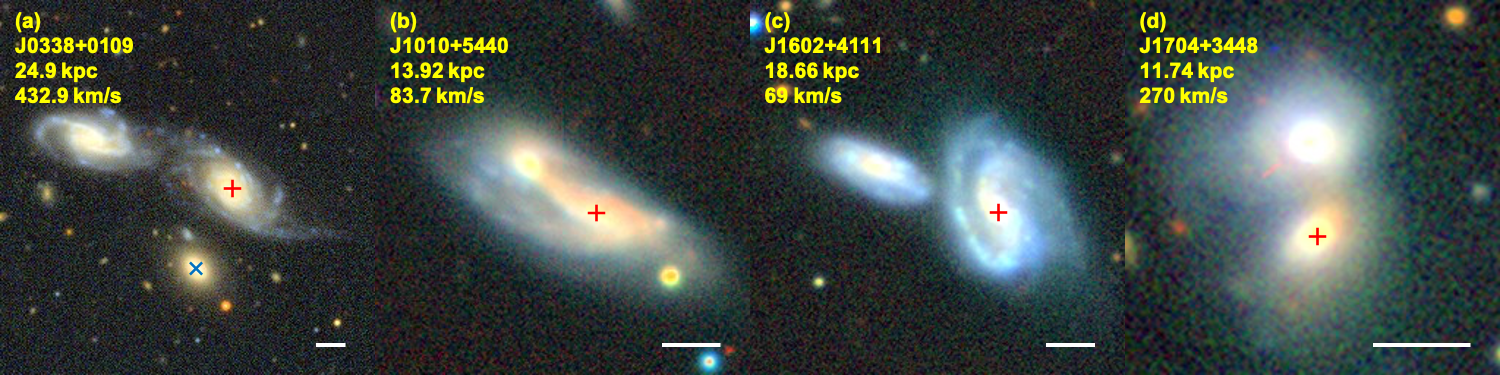}
\caption{DESI images of four classical bulge galaxies with strong star formation enhancement. The red plus signs mark the strongly enhanced galaxies. The blue cross in the image of J0338+0109 marks the E component of the pair. The white dash on the right-bottom corner of each stamp illustrates the angular scale of $10\arcsec$. The pair separation and velocity difference are provided as well.
\label{fig:classical}}
\end{figure*}

Fig.~\ref{fig:pseudo} and Fig.~\ref{fig:classical} are optical ($u$, $g$, $r$) color images of the nine pseudobulge galaxies with strong enhancement and four classical-bulge galaxies with strong enhancement, respectively. These images are taken from the database of Dark Energy Spectroscopic Instrument (DESI\footnote{\url{https://www.legacysurvey.org/}}) which are deeper and of better angular resolutions than SDSS images and therefore can show the diffuse features more clearly. Four pseudobulge galaxies J07543194+1648214, J09155467+4419510, J09155552+4419580, and J17045097+3449020 (panel b, c, g of Figure~\ref{fig:pseudo}) and two classical-bulge galaxies J10100079+5440198 and J17045089+3448530 (panel b, d of Figure~\ref{fig:classical}) are found in pairs seemed to be almost coalesced, namely in a late merging stage. This result suggests that a high fraction ($6/13 = 46\%$) of galaxies with strong enhancement are in late-stage mergers. Here we separate late-stage mergers (pairs with strong interaction signs and very close nuclei) from early-stage mergers (pairs without strong interaction signs or the two galaxies being widely separated). The late-stage mergers are likely seen during or after the second close encounter, because it takes relatively long time (more than a few 100 Myrs) for galaxies to develop the interaction features after the first close encounter. In general, only a low fraction of paired galaxies selected in optical and/or near-IR bands (such as H-KPAIR) are late-stage mergers because this stage lasts significantly shorter than the stage before it \citep{2007A&A...468...61D}. Also, for the H-KPAIR sample, the selection criterion of separation $\rm \geq 5\; h^{-1}\; kpc$ excludes some late-stage mergers, particularly those which are already coalesced. Hence, the fact that in our sample late-stage mergers represent 46\% of SFGs with strong enhancement suggests that the majority of SFGs in late-stage mergers may harbor strong tidally induced star formation whereas only a very low fraction of SFGs in early-stage mergers may do so. For the latter, significant SFR enhancement may be triggered by strong tidal torques only in SFGs that are undergoing or passed the first close encounter (namely pericenter passage in \citealt{2020ApJ...892L..20F}) in low-speed coplanar orbits (namely with well spin–orbit alignment in \citet{2021ApJ...909...34M}) as argued by \citep{2021ApJ...918...55X}.

The remaining seven galaxies with strong enhancement are in earlier-stage mergers with the two component galaxies clearly separated. Among them, J01183414-0013417, J13153506+6207286, and J16024257+4111501 (panel a, e of Figure~\ref{fig:pseudo} and panel c of Figure~\ref{fig:classical}) are likely in interaction systems with low-inclination orbits and low orbital velocity (all having $\rm \delta v < 70\; km\; s^{-1}$ where $\rm \delta v$ is the measured relative radial velocity of the companion). Particularly J13153506+6207286 (panel e of Figure~\ref{fig:pseudo}), which is in Arp~238 ($\rm \delta v = 6\; km\; s^{-1}$), has been confirmed to have a coplanar orbit by a simulation of \citet{2016MNRAS.459..720H}. \citet{2021ApJ...918...55X} found a very compact molecular gas concentration in its nucleus which is fueling a strong active nuclear starburst. They argued that the nuclear starburst is triggered by a strong tidal torque which has been predicted by simulations for galaxies in low-speed coplanar interactions \citep{1996ApJ...471..115B,2009ApJ...691.1168H}. This interpretation may also apply to J01183414-0013417 and J16024257+4111501. In contrast, J10332972+4404342, J14005783+4251203, and J14005879+4250427 (panel d, f of Figure~\ref{fig:pseudo}) may be in high-inclination orbits, and the latter two are members of the same pair (KPAIR J1400+4251) which has a high relative radial velocity ($\rm \delta v = 234\; km\; s^{-1}$). These three galaxies all have pseudobulges and might be bar galaxies from the appearance though it is uncertain given the limited resolution of the images. If this is true then their central starbursts may be field by the bars. Finally, J03381222+0110088 (panel a of Figure~\ref{fig:classical}) is the only one that is in an S+E pair among the 13 SFGs with strong enhancement. However, as shown in the optical image, it appears to be interacting with another nearby SFG which is in the same group of the S+E pair. The three galaxies form a close triplet dominated by SFGs. Therefore J03381222+0110088 is not in a de facto S+E pair. It has a very high total gas content with $\rm M_{H2+HI}/M_{star}= 0.56$ and a relatively low star formation efficiency of $\rm SFE=SFR/M_{H2}= 10^{-9.10}\; yr^{-1}$, which is below the average SFE of the H-KPAIR sample in \citet{2019A&A...627A.107L}. It appears that the high sSFR of J03381222+0110088 is 
mainly due to its very high gas content which may not be related to any tidal effect.

\subsection{Pseudo-bulge galaxies in S+S and S+E pairs} \label{subsec:ss-se}

In their study of the B/T dependence of $\rm sSFR_{enh}$, \citet{2022ApJS..261...34H} found that except for in the last bin of $\rm 0.5 < B/T \leq 1$ where SFGs in S+S and S+E pairs have similar means of $\rm sSFR_{enh}$ (both consistent with $\rm sSFR_{enh} = 0$), in all other B/T bins the former have higher $\rm sSFR_{enh}$ means than the latter. Particularly in the lowest bin of $\rm B/T \leq 0.1$, the difference is $\rm 0.4\pm0.1\;dex$. In \citet{2022ApJS..261...34H} pseudobulge galaxies are assigned with the same $\rm B/T=0$ and therefore are included in the bin $\rm B/T \leq 0.1$. Do pseudobulge SFGs in S+S and in S+E pairs have significantly different $\rm sSFR_{enh}$?

\begin{figure}[htb!]
\includegraphics[width=\hsize]{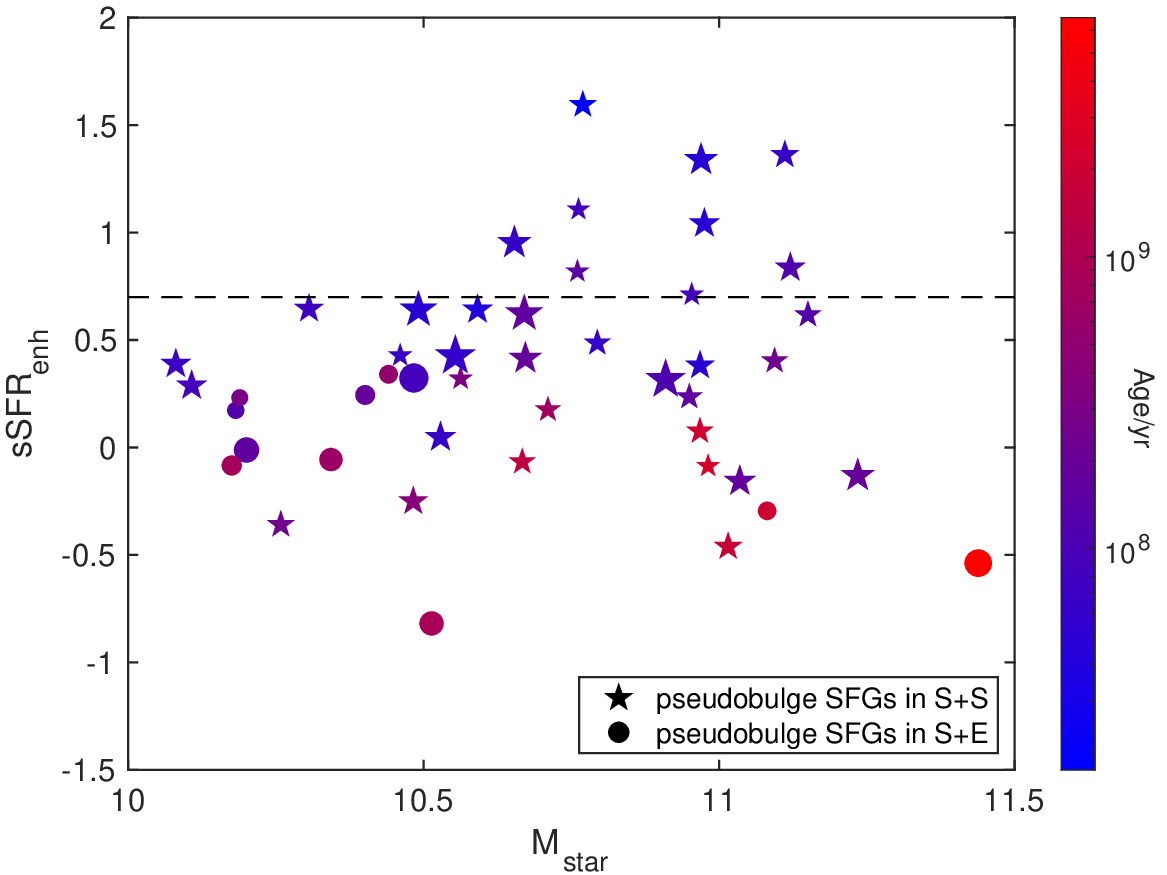}
\caption{Plot of $\rm sSFR_{enh}$ vs.~$\rm M_{star}$ for pseudo-bulge SFGs in S+S pairs (pentagram) and in S+E pairs (circle). The symbol color indicates the Age of the central stellar population, and the symbol size shows the B/T ratio. The dashed line corresponds to $\rm sSFR_{enh}=0.7$.
\label{fig:enh-m}}
\end{figure}

In Fig.~\ref{fig:enh-m} we compare pseudobulge SFGs in S+S pairs and in S+E pairs. It plots $\rm sSFR_{enh}$ against the stellar mass $\rm M_{star}$, with the color scale indicating the Age and the symbol size the B/T ratio. The plot shows clearly that, compared to pseudobulge SFGs in S+S pairs, those in S+E pairs tend to have older stellar populations and lower sSFR enhancements and none has strong enhancement ($\rm sSFR_{enh} \geq 0.7$). The means of $\rm sSFR_{enh}$ are $0.45\pm0.08$ and $-0.04\pm0.11$ for pseudobulge SFGs in S+S pairs and in S+E pairs, respectively, and the difference is above $3\;\sigma$. Most of the pseudobulge SFGs in S+E pairs (9 of 11) have relatively low stellar mass ($\leq 10^{10.6}\;M_\odot$).

\begin{figure}[htb!]
\includegraphics[width=\hsize]{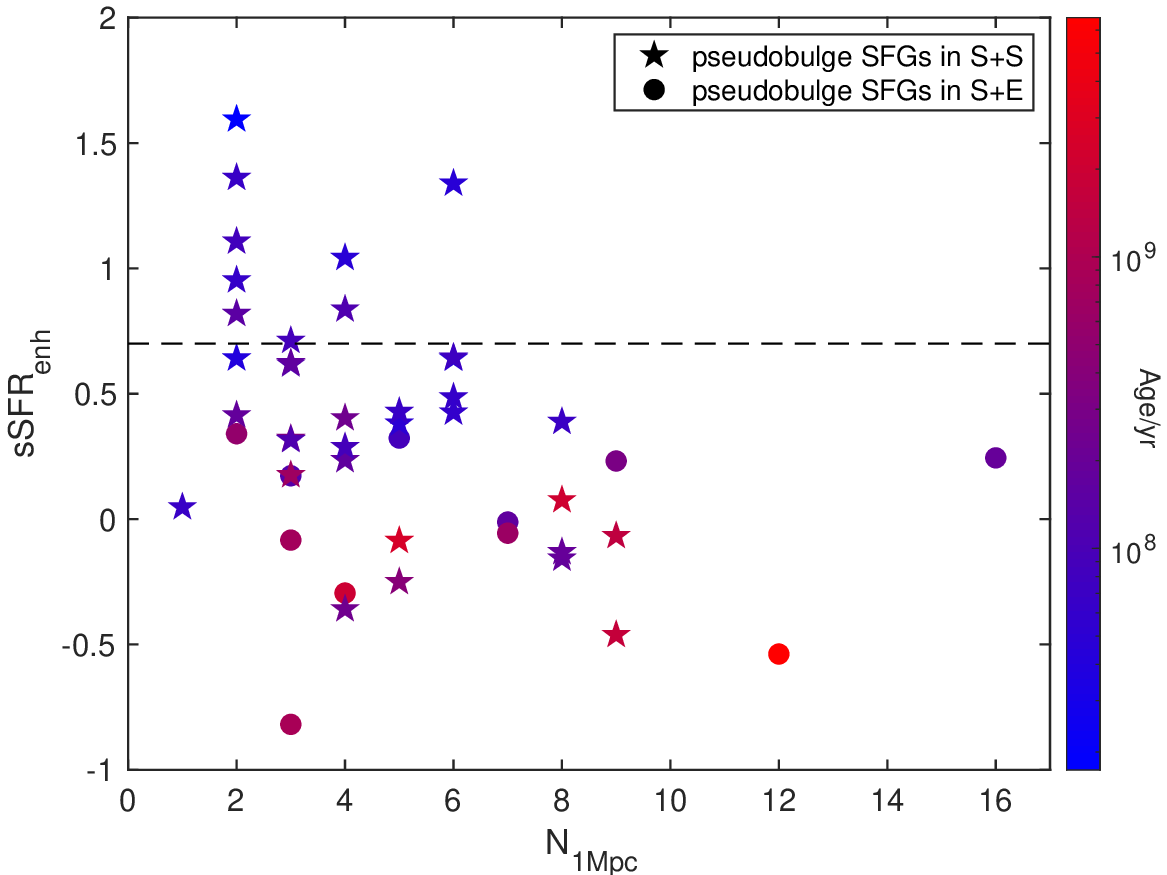}
\caption{Plot of $\rm sSFR_{enh}$ vs.~$\rm N_{1Mpc}$, where $\rm N_{1Mpc}$ is an indicator of local density defined by the number of galaxies brighter than $\rm M_{r} = -19.7\; mag$ within 1~Mpc projected distance around the galaxy in question \citep{2022ApJS..261...34H}. filled pentagrams and filled circles represent pseudobulge SFGs in S+S pairs and in S+E pairs, respectively. The dashed line corresponds to $\rm sSFR_{enh}=0.7$. 
\label{fig:enh-dens}}
\end{figure}

\begin{figure}[htb!]
\includegraphics[width=\hsize]{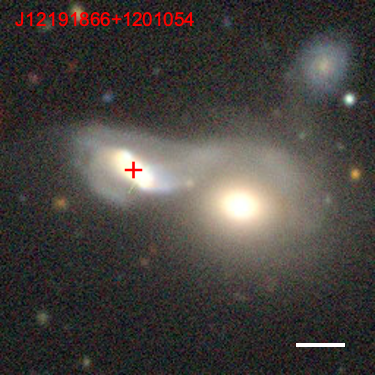}
\caption{DESI images of J1219+1200, an S+E pair. S component, J12191866+1201054 (identified by a red plus sign), is a pseudobulge SFG that contains a bar. The white dash on the right-bottom corner of the figure illustrates the angular scale of $10\arcsec$.
\label{fig:SE-bar}}
\end{figure}

Fig.~\ref{fig:enh-dens} compares pseudobulge SFGs in S+S and S+E in a plot of $\rm sSFR_{enh}$ versus the local density indicator $\rm N_{1Mpc}$ (defined in \cite{2022ApJS..261...34H} and also introduced in Sect.~\ref{subsec:pseudo-classical}). SFGs in high-density regions with $\rm N_{1Mpc} \geq 7$ have significantly lower $\rm sSFR_{enh}$ than the rest of the sample, and a higher fraction (5/11) of pseudobulge SFGs in S+E, compared to those in S+S, are in this category. These SFGs should be in rich groups/clusters, where star formation may be quenched by galaxy harassment \citep{1998ApJ...495..139M} and/or ram-pressure stripping \citep{1972ApJ...176....1G,1983AJ.....88..881G,2006A&A...446..839G}. Also, these paired SFGs may have very low probability of being in a coplanar orbit because of the frequent disturbance and therefore are less likely to have nuclear starbursts \citep{2021ApJ...918...55X}. Detailed inspections of the optical images show that their pseudobulges often correspond to bars or inner rings with low star formation activity (for example J12191866+1201054 which has $\rm sSFR_{enh} = -0.01$, Fig.~\ref{fig:SE-bar}).

\begin{figure}[htb!]
\includegraphics[width=\hsize]{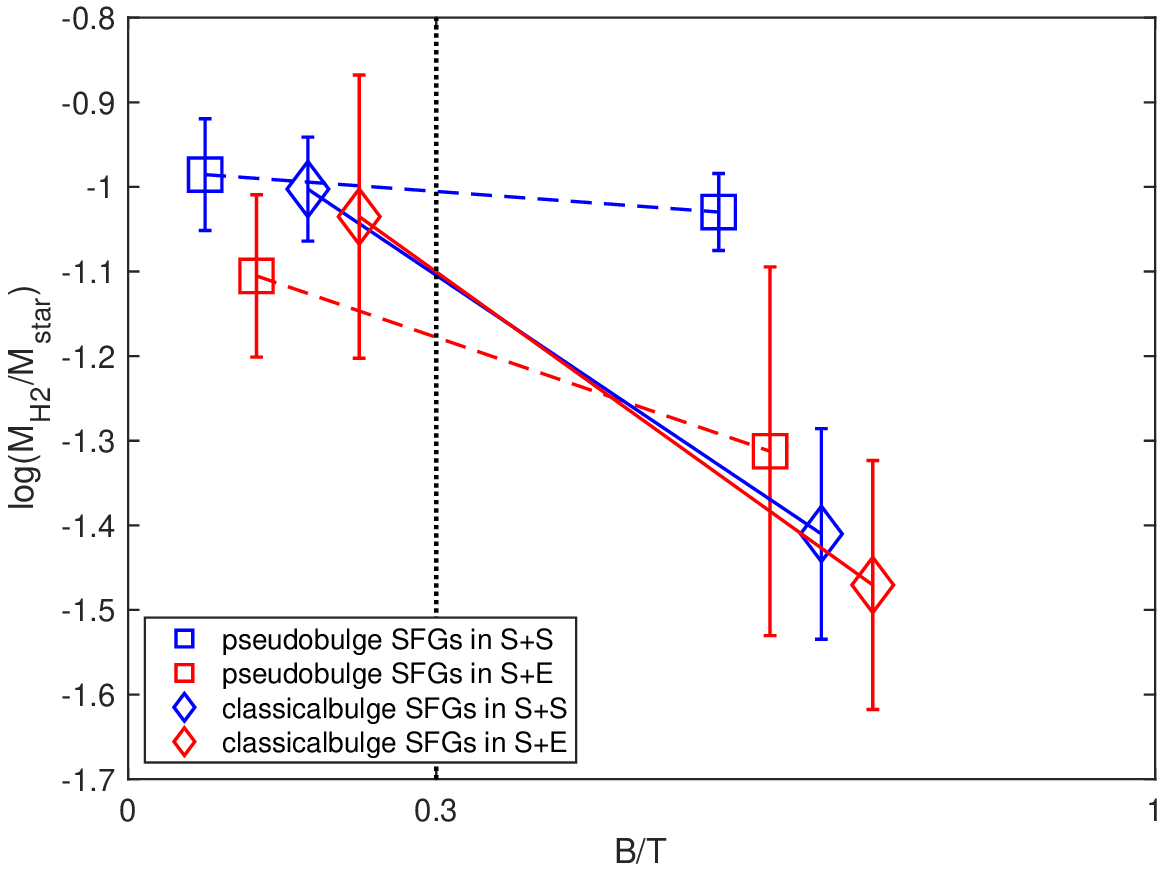}
\caption{Plot of means of $\rm \log (M_{H2}/M{star})$ and errors in two B/T bins for pseudobulge SFGs (square) and classical bulge SFGs (diamond) in S+S (blue) and S+E (red) pair.
\label{fig:m-h2}}
\end{figure}

\begin{figure}[htb!]
\includegraphics[width=\hsize]{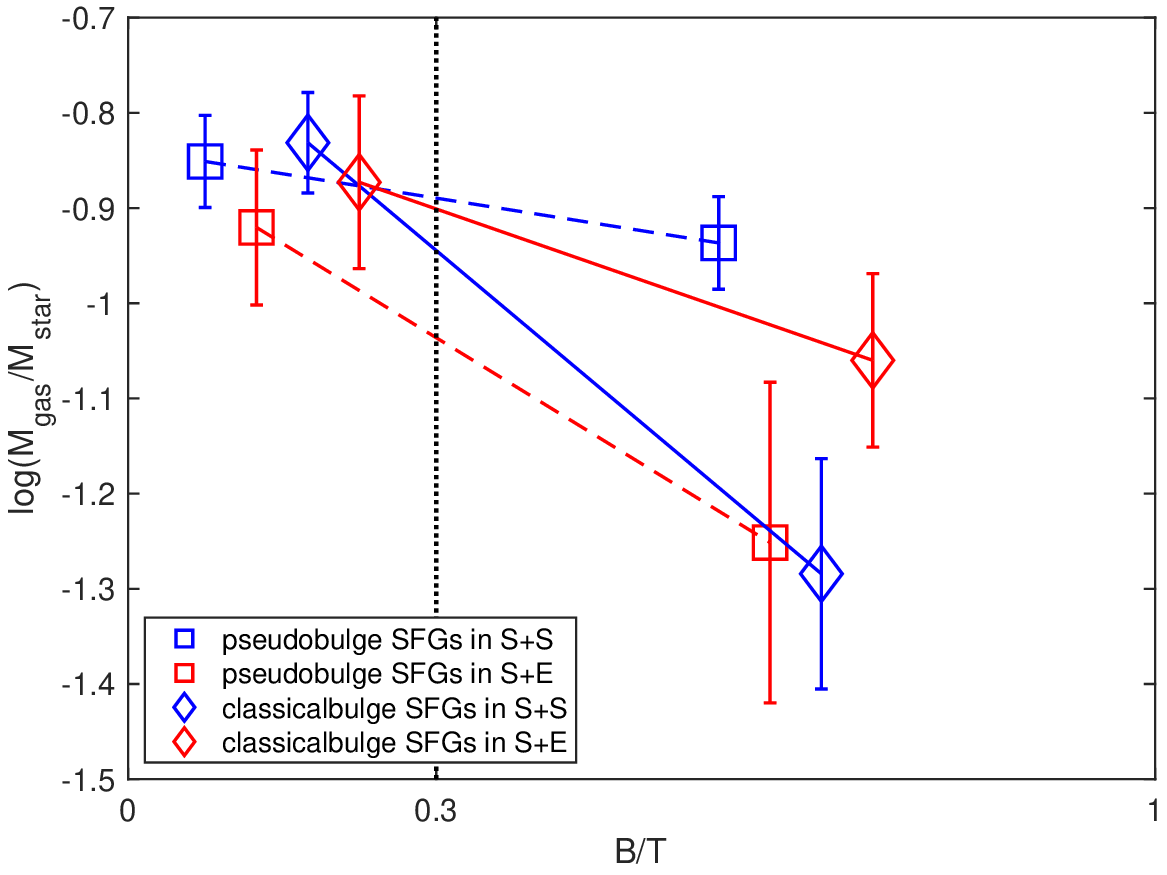}
\caption{Plot of means of $\rm \log (M_{gas}/M{star})$ and errors in two B/T bins for pseudobulge SFGs (square) and classical bulge SFGs (diamond) in S+S (blue) and S+E (red) pair.
\label{fig:m-gas}}
\end{figure}

\begin{figure}[htb!]
\includegraphics[width=\hsize]{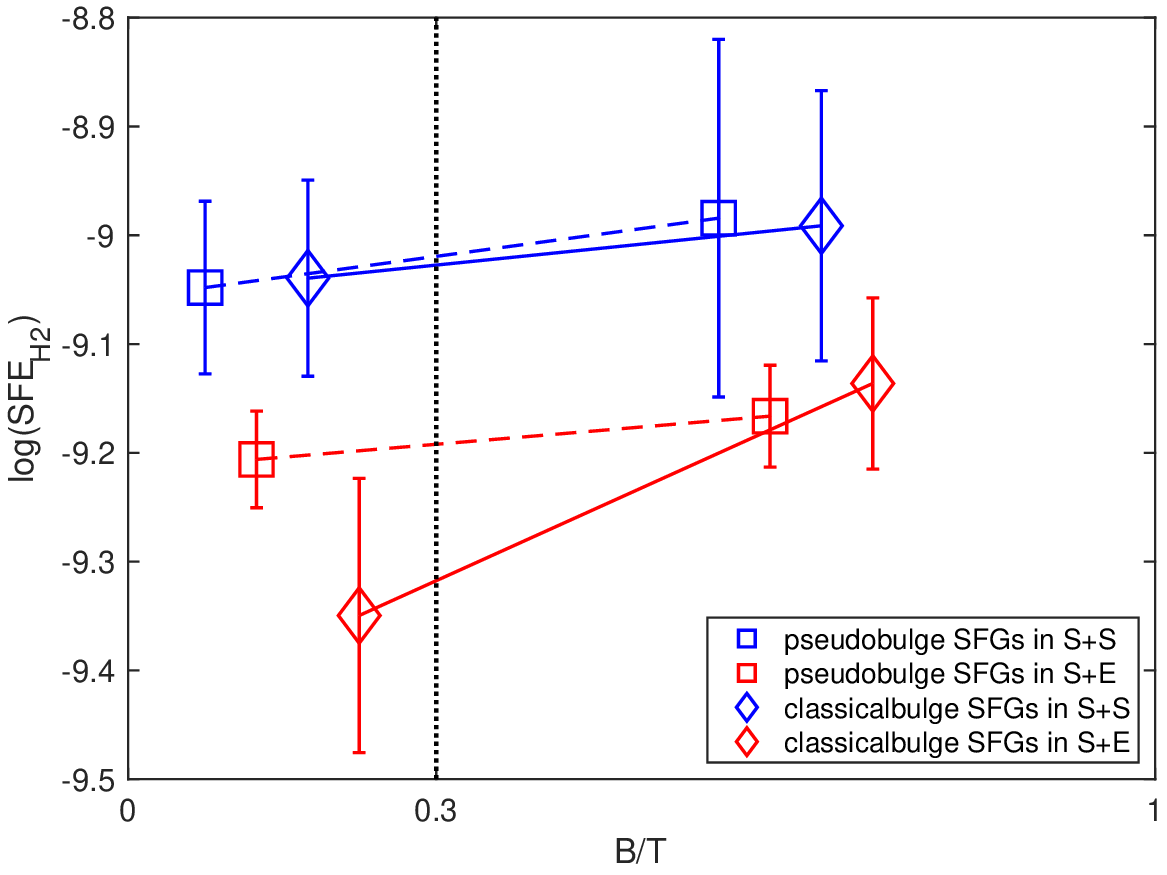}
\caption{Plot of means of $\rm \log (SFE_{H2})$ and errors in two B/T bins for pseudobulge SFGs (square) and classical bulge SFGs (diamond) in S+S (blue) and S+E (red) pair.
\label{fig:sfe-h2}}
\end{figure}

\begin{figure}[htb!]
\includegraphics[width=\hsize]{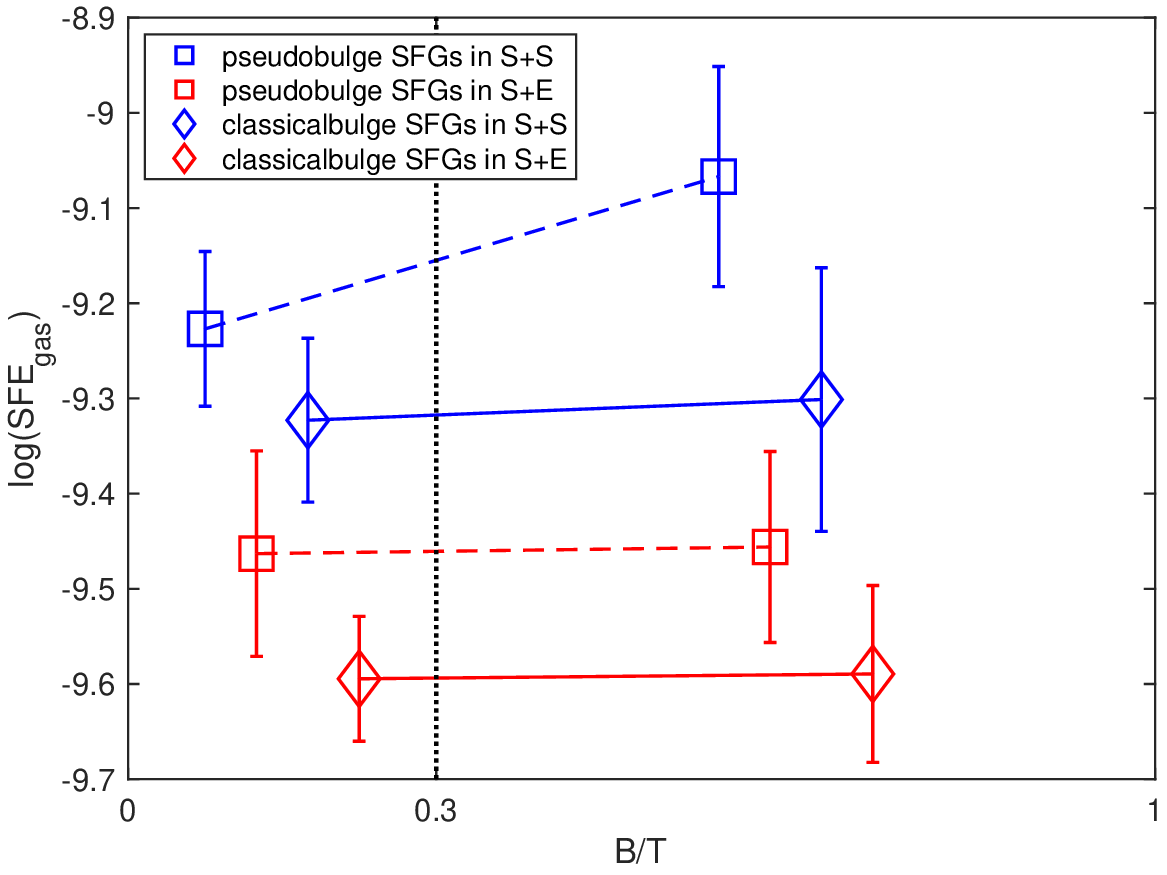}
\caption{Plot of means of $\rm \log (SFE_{gas})$ and errors in two B/T bins for pseudobulge SFGs (square) and classical bulge SFGs (diamond) in S+S (blue) and S+E (red) pair.
\label{fig:sfe-gas}}
\end{figure}

The mean and errors of H$_2$ gas mass fraction\citep{2019A&A...627A.107L} and that of the total gas mass derived from FIR dust mass \citep{2016ApJS..222...16C} were plotted in two B/T bins: (1) disk dominant ($\rm B/T \leq 0.3$), (2) bulge dominant ($\rm B/T>0.3$) in Figure~\ref{fig:m-h2} and~\ref{fig:m-gas}, as well the SFE calculated from them in Figure~\ref{fig:sfe-h2} and~\ref{fig:sfe-gas}, separately. Generally, there is a decreasing sign for both H$_2$ gas and the total gas on B/T. The pseudobulge SFGs in S+S pairs have the most flat slope, while they have the highest gas fractions in bulge dominant galaxy (bin 2). No obvious trend is found in either SFE versus B/T. When comparing the subsamples, the pseudobulge SFGs have higher $\rm SFE_{gas}$, and the SFGs in S+S pairs take the lead in both SFE. The most attractive point is that the pseudobulge galaxies in S+E pairs have not as high SFE as the classical bulge galaxies in S+S pairs. 

\section{Discussion} \label{sect:discussion}
The significant sSFR enhancement of close major-merger pairs is 
mainly due to a small population of SFGs with strong enhancement ($\sim 15\%$ in our sample; Fig.~\ref{fig:enh-BT} ). About half of these SFGs are found in late-stage mergers, all of which are in S+S pairs and none in an S+E pair. It is unlikely that any of the late-stage S+S mergers containing strong enhanced SFGs can be misclassified S+E mergers because none of the galaxies in these pairs is dominated by a classical bulge (the largest classical bulge is found in J17045088+3448529 shown in panel d of Figure~\ref{fig:classical} with a $\rm B/T=0.48$).

\begin{figure*}[htb!]
\includegraphics[width=\hsize]{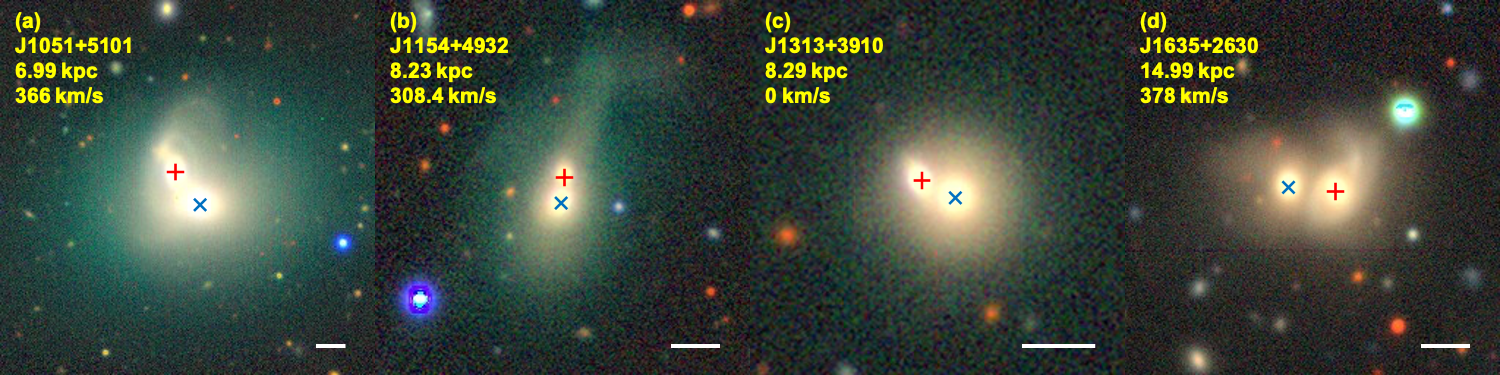}
\caption{DESI images of four late-stage S+E mergers. The red plus signs mark the S components and the blue crosses the E components. The white dash on the right-bottom corner of each stamp illustrates the angular scale of $10\arcsec$. The pair separation and velocity difference are provided as well.
\label{fig:merger-SE}}
\end{figure*}

What is the reason for spirals in late-stage S+E mergers to avoid strong starbursts? There are four S+E late-stage mergers in H-KPAIR, which are listed in Table~\ref{tab:4}. Each of them has the two member galaxies nearly coalesced within a common halo (Fig.~\ref{fig:merger-SE}). They are pairs of massive galaxies with the primaries more massive than $10^{11}/M_{\odot}$. All spiral galaxies in these pairs are rather quiescent in star formation and only one of them (J16354293+2630494; panel d) is included in our SFG sample with an $\rm sSFR =10^{-11.26}\;yr^{-1}$. They are uniformly gas-poor with the $\rm M_{gas}/M_{star}$ ratio less than 4\% \citep{2016ApJS..222...16C}. which is lower than the ratios of 88 percent of SFGs in the H-KPAIR sample, and significantly below the mean of the sample ($\rm M_{gas}/M_{star}= 0.13\pm0.01$). 

Three (J10514450+5101303, J13131470+3910382, and J16354293+2630494) are in pairs containing the most massive galaxies of rich groups or clusters that have more than 12 members \citep{2012ApJ...752...41Y}. For them, the rich-group/cluster environment may be the key to the lack of cold gas and star formation because the ram-pressure stripping and/or evaporation by the hot gas in IGM can remove most of the cold gas in the ISM and tidal tails. J11542299+4932509 is apparently in an isolated pair. Its companion (J11542307+4932456) is a very massive Elliptical with $\rm M_{star} = 10^{11.35}\;M_\odot$. The pair is detected by ROSAT in X-ray with a flux of $\rm f_{0.2-2\;keV} = 8.22 (\pm1.73) \times 10^{-14}\;erg\;s^{-1}\;cm^{-2}$. This may be indicative of the presence of a hot gas halo around the pair which can strip/evaporate the cold gas and be responsible for the low gas content ($\rm M_{gas}/M_{star} = 0.035$) and low sSFR ($< 10^{-11.33}\;yr^{-1}$) of J11542299+4932509. 

Thus, it appears that the lack of strong enhancement in late-stage S+E mergers may be due to the removal of cold gas by hot gas in the immediate environment \citep{2009ApJ...691.1828P,2011A&A...535A..60H}. It should be pointed out that the quiescent spiral galaxies in these late-stage S+E pairs are different from the SFGs studied by \citet{2016ApJS..222...16C} and \citet{2019A&A...627A.107L}. These authors found no significant difference in $\rm M_{gas}/M_{star}$ and in $\rm M_{H2+HI}/M_{star}$ between SFGs in S+S and S+E pairs, and therefore refuted the ``cold-gas stripping'' hypothesis as far as SFGs in S+E pairs are concerned. A majority of spirals in S+E pairs are SFGs (64\% =28/44 for H-KPAIR), and many of them are gas-rich (e.g. NGC~2936 in Arp~142; \citealt{2021ApJ...918...55X}). It will be interesting to find out why they are absent in late-stage S+E mergers.
 
The interaction-induced star formation enhancement is affected by many factors, some seem to be necessary conditions (e.g., a galaxy should have gas to fuel star formation), but it's hard to say which is a sufficient condition. For instance, the four most enhanced classical bulge SFGs (Figure~\ref{fig:classical}) never reached the ultra enhanced level (say, $\rm sSFR_{enh}>1$, corresponding to more than 10 times sSFR enhancement compared with the median of the matching controls) as their pseudobulge counterparts. However, as the outstanding cases in classical SFGs, three of them have very low B/T. The other one J17045088+3448529 with $\rm B/T=0.48$, as well the ``blue'' bulge dominated galaxy J13151726+4424255 whose $\rm sSFR_{enh}$ reaches almost to 0.7 as we talked about the Figure~\ref{fig:age-BT} (also see the most upper-right data point in Figure~\ref{fig:enh-BT}), are located in low density environment ($\rm N_{1Mpc}$ equal to 4 and 3, respectively). 

Though KPAIR \citep{2009ApJ...695.1559D} was selected from allsky, strict selection criteria make it a relatively small sample. After further refining the classifications such as separating the sample into S+S/S+E pair or pseudo/classical bulge, the subsamples sometimes are too small for a very robust for a robust statistical analysis. Nevertheless, the individual case studies in this paper provide a perspective of the complex mechanism of the interaction-induced star formation enhancement. Future sky surveys with deeper sensitivities and sharper resolutions (e.g. Euclid) will produce larger and better samples for more conclusive studies. 

\section{Conclusion} \label{sect:conclusion}
Here we present a study of SFGs with pseudobulges (bulges with S\'ersic index n $< 2$) in a local close major-merger galaxy pair sample (H-KPAIR). The sample is taken from \citet{2022ApJS..261...34H} who did 2-D {\sc galfit} decompositions to study the B/T dependence of the star formation enhancement of paired SFGs. New aperture photometries in the SDSS $u$, $r$, $i$-bands and the 2MASS K$_s$-band are carried out for the sample to investigate the stellar populations in the central part ($\rm D=7\; kpc$) of the galaxies. With the new data and data from the literature, we obtain the following results:
\begin{itemize}
\item The mean Age of central stellar populations in spiral galaxies with classical bulges increases with the B/T ratio. On the other hand, the mean Age of central stellar populations in Spirals with pseudobulges is nearly constant against the B/T ratio and is consistent with the mean Age of central stellar populations of disky galaxies (with $\rm B/T \sim 0$). This confirms that the pseudobulges in the sample are associated mainly with disky phenomena such as bars, nuclear rings, and bright nuclei. 
\item Spiral galaxies have a modestly reduced fraction of pure disk galaxies than in single Spirals, which can be due to the misidentifying as a large pseudobulge of the nuclear/circum-nuclear starburst phenomena.
\item Compared with SFGs with classical bulges, those with pseudobulges have a $2\;\sigma$ higher mean of sSFR enhancement ($\rm sSFR_{enh} = 0.33\pm0.07$ vs $\rm sSFR_{enh}= 0.12\pm0.06$) and much broader scatter ($\sim 1$\;dex).
\item The eight SFGs that have the highest $\rm sSFR_{enh}$ in the sample are all pseudobulge galaxies. And a majority (69\%) of paired SFGs with strong enhancement (having sSFR more than 5 times the median of the control galaxies) have pseudobulges. The Spitzer data show that the pseudobulges in these galaxies are tightly linked to nuclear starbursts. These results suggest that nuclear starburst may be the dominant mode for strong tidally induced star formation in paired galaxies.
\item Pseudobulge SFGs in S+S and in S+E pairs have $3\;\sigma$ significantly different sSFR enhancement: the mean $\rm sSFR_{enh} = 0.45\pm0.08$ and $-0.04\pm0.11$ for the former and latter, respectively. A high fraction (5/11) of pseudobulge SFGs in S+E are in rich groups/clusters (local density $\rm N_{1Mpc} \geq 7$), whose star formation enhancement may be hindered by the environment.
\end{itemize}
In conclusion, we find that paired SFGs with pseudobulges have very diversified central stellar populations and sSFR enhancements. They include galaxies with bars, inner disks and rings, and strong nuclear/circum-nuclear starbursts. The SFGs with strong sSFR enhancement are dominated by pseudobulge galaxies with nuclear/circum-nuclear starbursts. On the other hand, many pseudobulge SFGs, particularly those in S+E pairs, are barred or ringed galaxies with old central stellar populations and low sSFR enhancements, which are probably caused by environmental effects.

\begin{acknowledgements}
This work is supported by National Natural Science Foundation of China (NSFC) No. 11873055 and No.11933003, and is sponsored (in part) by the Chinese Academy of Sciences (CAS) through a grant to the CAS South America Center for Astronomy (CASSACA). UL acknowledges support from project PID2020-114414GB-100, financed by MCIN/AEI/10.13039/501100011033 and from the Junta de Andalucia\'ia (Spain) grant FQM108. TF and QY acknowledge support by National Key R\&D Program of China No. 2017YFA0402600, National Natural Science Foundation of China (NSFC) grant Nos. 11890692, 12133008, and 12221003 and China Manned Space Project No. CMS-CSST-2021-A04.
\\
The authors would like to thank Gaoxiang Jin, Cheng Cheng, Chen Cao, Yilun Wang, and Yufu Shen for their helpful discussions.
\\
This publication makes use of data products from the Sloan Digital Sky Survey (SDSS). Funding for the Sloan Digital Sky Survey IV has been provided by the Alfred P. Sloan Foundation, the U.S. Department of Energy Office of Science, and the Participating Institutions. SDSS acknowledges support and resources from the Center for High-Performance Computing at the University of Utah. The SDSS website is \url{www.sdss.org}. SDSS is managed by the Astrophysical Research Consortium for the Participating Institutions of the SDSS Collaboration including the Brazilian Participation Group, the Carnegie Institution for Science, Carnegie Mellon University, Center for Astrophysics | Harvard \& Smithsonian (CfA), the Chilean Participation Group, the French Participation Group, Instituto de Astrofísica de Canarias, The Johns Hopkins University, Kavli Institute for the Physics and Mathematics of the Universe (IPMU) / University of Tokyo, the Korean Participation Group, Lawrence Berkeley National Laboratory, Leibniz Institut für Astrophysik Potsdam (AIP), Max-Planck-Institut für Astronomie (MPIA Heidelberg), Max-Planck-Institut für Astrophysik (MPA Garching), Max-Planck-Institut für Extraterrestrische Physik (MPE), National Astronomical Observatories of China, New Mexico State University, New York University, University of Notre Dame, Observatório Nacional / MCTI, The Ohio State University, Pennsylvania State University, Shanghai Astronomical Observatory, United Kingdom Participation Group, Universidad Nacional Autónoma de México, University of Arizona, University of Colorado Boulder, University of Oxford, University of Portsmouth, University of Utah, University of Virginia, University of Washington, University of Wisconsin, Vanderbilt University, and Yale University.
\\
This publication makes use of data products from the Two Micron All Sky Survey, which is a joint project of the University of Massachusetts and the Infrared Processing and Analysis Center/California Institute of Technology, funded by the National Aeronautics and Space Administration and the National Science Foundation.
\\

This publication makes use of image products from the Dark Energy Spectroscopic Instrument (DESI). DESI construction and operations is managed by the Lawrence Berkeley National Laboratory. This research is supported by the U.S. Department of Energy, Office of Science, Office of High-Energy Physics, under Contract No. DE–AC02–05CH11231, and by the National Energy Research Scientific Computing Center, a DOE Office of Science User Facility under the same contract. Additional support for DESI is provided by the U.S. National Science Foundation, Division of Astronomical Sciences under Contract No. AST-0950945 to the NSF’s National Optical-Infrared Astronomy Research Laboratory; the Science and Technologies Facilities Council of the United Kingdom; the Gordon and Betty Moore Foundation; the Heising-Simons Foundation; the French Alternative Energies and Atomic Energy Commission (CEA); the National Council of Science and Technology of Mexico (CONACYT); the Ministry of Science and Innovation of Spain, and by the DESI Member Institutions. The DESI collaboration is honored to be permitted to conduct astronomical research on Iolkam Du’ag (Kitt Peak), a mountain with particular significance to the Tohono O'odham Nation.
\end{acknowledgements}

\bibliographystyle{raa} 
\bibliography{sample631} 

\begin{thebibliography}{61}
\providecommand\natexlab[1]{#1}
\providecommand\JournalTitle[1]{#1}

\bibitem[{Alonso} {et~al.}(2004)]{2004MNRAS.352.1081A}
{Alonso}, M.~S., {Tissera}, P.~B., {Coldwell}, G., \& {Lambas}, D.~G. 2004, \mnras, 352, 1081

\bibitem[{Barnes} \& {Hernquist}(1996)]{1996ApJ...471..115B}
{Barnes}, J.~E., \& {Hernquist}, L. 1996, \apj, 471, 115

\bibitem[{Bertin} {et~al.}(2002)]{2002ASPC..281..228B}
{Bertin}, E., {Mellier}, Y., {Radovich}, M., {et~al.} 2002, in Astronomical Society of the Pacific Conference Series, Vol. 281, Astronomical Data Analysis Software and Systems XI, ed. D.~A. {Bohlender}, D.~{Durand}, \& T.~H. {Handley}, 228

\bibitem[{Bluck} {et~al.}(2014)]{2014MNRAS.441..599B}
{Bluck}, A. F.~L., {Mendel}, J.~T., {Ellison}, S.~L., {et~al.} 2014, \mnras, 441, 599

\bibitem[{Bruzual} \& {Charlot}(2003)]{2003MNRAS.344.1000B}
{Bruzual}, G., \& {Charlot}, S. 2003, \mnras, 344, 1000

\bibitem[{Cao} {et~al.}(2016)]{2016ApJS..222...16C}
{Cao}, C., {Xu}, C.~K., {Domingue}, D., {et~al.} 2016, \apjs, 222, 16

\bibitem[{Cardelli} {et~al.}(1989)]{1989ApJ...345..245C}
{Cardelli}, J.~A., {Clayton}, G.~C., \& {Mathis}, J.~S. 1989, \apj, 345, 245

\bibitem[{Chown} {et~al.}(2019)]{2019MNRAS.484.5192C}
{Chown}, R., {Li}, C., {Athanassoula}, E., {et~al.} 2019, \mnras, 484, 5192

\bibitem[{Cox} {et~al.}(2008)]{2008MNRAS.384..386C}
{Cox}, T.~J., {Jonsson}, P., {Somerville}, R.~S., {Primack}, J.~R., \& {Dekel}, A. 2008, \mnras, 384, 386

\bibitem[{Di Matteo} {et~al.}(2008)]{2008A&A...492...31D}
{Di Matteo}, P., {Bournaud}, F., {Martig}, M., {et~al.} 2008, \aap, 492, 31

\bibitem[{Di Matteo} {et~al.}(2007)]{2007A&A...468...61D}
{Di Matteo}, P., {Combes}, F., {Melchior}, A.~L., \& {Semelin}, B. 2007, \aap, 468, 61

\bibitem[{Domingue} {et~al.}(2009)]{2009ApJ...695.1559D}
{Domingue}, D.~L., {Xu}, C.~K., {Jarrett}, T.~H., \& {Cheng}, Y. 2009, \apj, 695, 1559

\bibitem[{Driver} {et~al.}(2006)]{2006MNRAS.368..414D}
{Driver}, S.~P., {Allen}, P.~D., {Graham}, A.~W., {et~al.} 2006, \mnras, 368, 414

\bibitem[{Ellison} {et~al.}(2010)]{2010MNRAS.407.1514E}
{Ellison}, S.~L., {Patton}, D.~R., {Simard}, L., {et~al.} 2010, \mnras, 407, 1514

\bibitem[{Erwin} {et~al.}(2021)]{2021MNRAS.502.2446E}
{Erwin}, P., {Seth}, A., {Debattista}, V.~P., {et~al.} 2021, \mnras, 502, 2446

\bibitem[{Feng} {et~al.}(2020)]{2020ApJ...892L..20F}
{Feng}, S., {Shen}, S.-Y., {Yuan}, F.-T., {Riffel}, R.~A., \& {Pan}, K. 2020, \apjl, 892, L20

\bibitem[{Fraser-McKelvie} {et~al.}(2020)]{2020MNRAS.499.1116F}
{Fraser-McKelvie}, A., {Merrifield}, M., {Arag{\'o}n-Salamanca}, A., {et~al.} 2020, \mnras, 499, 1116

\bibitem[{Gavazzi} {et~al.}(2006)]{2006A&A...446..839G}
{Gavazzi}, G., {Boselli}, A., {Cortese}, L., {et~al.} 2006, \aap, 446, 839

\bibitem[{Giovanelli} \& {Haynes}(1983)]{1983AJ.....88..881G}
{Giovanelli}, R., \& {Haynes}, M.~P. 1983, \aj, 88, 881

\bibitem[{Gunn} \& {Gott}(1972)]{1972ApJ...176....1G}
{Gunn}, J.~E., \& {Gott}, J.~Richard, I. 1972, \apj, 176, 1

\bibitem[{He} {et~al.}(2022)]{2022ApJS..261...34H}
{He}, C., {Xu}, C.~K., {Domingue}, D., {Cao}, C., \& {Huang}, J.-s. 2022, \apjs, 261, 34

\bibitem[{Herrera-Endoqui} {et~al.}(2017)]{2017A&A...599A..43H}
{Herrera-Endoqui}, M., {Salo}, H., {Laurikainen}, E., \& {Knapen}, J.~H. 2017, \aap, 599, A43

\bibitem[{Holincheck} {et~al.}(2016)]{2016MNRAS.459..720H}
{Holincheck}, A.~J., {Wallin}, J.~F., {Borne}, K., {et~al.} 2016, \mnras, 459, 720

\bibitem[{Hopkins} {et~al.}(2009)]{2009ApJ...691.1168H}
{Hopkins}, P.~F., {Cox}, T.~J., {Younger}, J.~D., \& {Hernquist}, L. 2009, \apj, 691, 1168

\bibitem[{Hwang} {et~al.}(2010)]{2010A&A...522A..33H}
{Hwang}, H.~S., {Elbaz}, D., {Lee}, J.~C., {et~al.} 2010, \aap, 522, A33

\bibitem[{Hwang} {et~al.}(2011)]{2011A&A...535A..60H}
{Hwang}, H.~S., {Elbaz}, D., {Dickinson}, M., {et~al.} 2011, \aap, 535, A60

\bibitem[{Keel} {et~al.}(1985)]{1985AJ.....90..708K}
{Keel}, W.~C., {Kennicutt}, R.~C., J., {Hummel}, E., \& {van der Hulst}, J.~M. 1985, \aj, 90, 708

\bibitem[{Kennicutt}(1998)]{1998ARA&A..36..189K}
{Kennicutt}, Robert~C., J. 1998, \araa, 36, 189

\bibitem[{Kennicutt} {et~al.}(1987)]{1987AJ.....93.1011K}
{Kennicutt}, Robert~C., J., {Keel}, W.~C., {van der Hulst}, J.~M., {Hummel}, E., \& {Roettiger}, K.~A. 1987, \aj, 93, 1011

\bibitem[{Kim} {et~al.}(2016)]{2016ApJS..225....6K}
{Kim}, K., {Oh}, S., {Jeong}, H., {et~al.} 2016, \apjs, 225, 6

\bibitem[{Kormendy} \& {Ho}(2013)]{2013ARA&A..51..511K}
{Kormendy}, J., \& {Ho}, L.~C. 2013, \araa, 51, 511

\bibitem[{Kormendy} \& {Kennicutt}(2004)]{2004ARA&A..42..603K}
{Kormendy}, J., \& {Kennicutt}, Robert~C., J. 2004, \araa, 42, 603

\bibitem[{Larson} \& {Tinsley}(1978)]{1978ApJ...219...46L}
{Larson}, R.~B., \& {Tinsley}, B.~M. 1978, \apj, 219, 46

\bibitem[{Li} {et~al.}(2023)]{2023ApJ...953...91L}
{Li}, Y.~A., {Ho}, L.~C., \& {Shangguan}, J. 2023, \apj, 953, 91

\bibitem[{Lintott} {et~al.}(2008)]{2008MNRAS.389.1179L}
{Lintott}, C.~J., {Schawinski}, K., {Slosar}, A., {et~al.} 2008, \mnras, 389, 1179

\bibitem[{Lisenfeld} {et~al.}(2019)]{2019A&A...627A.107L}
{Lisenfeld}, U., {Xu}, C.~K., {Gao}, Y., {et~al.} 2019, \aap, 627, A107

\bibitem[{Madau} \& {Dickinson}(2014)]{2014ARA&A..52..415M}
{Madau}, P., \& {Dickinson}, M. 2014, \araa, 52, 415

\bibitem[{Martig} {et~al.}(2009)]{2009ApJ...707..250M}
{Martig}, M., {Bournaud}, F., {Teyssier}, R., \& {Dekel}, A. 2009, \apj, 707, 250

\bibitem[{Meert} {et~al.}(2015)]{2015MNRAS.446.3943M}
{Meert}, A., {Vikram}, V., \& {Bernardi}, M. 2015, \mnras, 446, 3943

\bibitem[{Mihos} \& {Hernquist}(1996)]{1996ApJ...464..641M}
{Mihos}, J.~C., \& {Hernquist}, L. 1996, \apj, 464, 641

\bibitem[{Moon} {et~al.}(2019)]{2019ApJ...882...14M}
{Moon}, J.-S., {An}, S.-H., \& {Yoon}, S.-J. 2019, \apj, 882, 14

\bibitem[{Moon} {et~al.}(2021)]{2021ApJ...909...34M}
{Moon}, J.-S., {An}, S.-H., \& {Yoon}, S.-J. 2021, \apj, 909, 34

\bibitem[{Moore} {et~al.}(1998)]{1998ApJ...495..139M}
{Moore}, B., {Lake}, G., \& {Katz}, N. 1998, \apj, 495, 139

\bibitem[{Nikolic} {et~al.}(2004)]{2004MNRAS.355..874N}
{Nikolic}, B., {Cullen}, H., \& {Alexander}, P. 2004, \mnras, 355, 874

\bibitem[{Pan} {et~al.}(2019)]{2019ApJ...881..119P}
{Pan}, H.-A., {Lin}, L., {Hsieh}, B.-C., {et~al.} 2019, \apj, 881, 119

\bibitem[{Park} \& {Choi}(2009)]{2009ApJ...691.1828P}
{Park}, C., \& {Choi}, Y.-Y. 2009, \apj, 691, 1828

\bibitem[{Patton} {et~al.}(2013)]{2013MNRAS.433L..59P}
{Patton}, D.~R., {Torrey}, P., {Ellison}, S.~L., {Mendel}, J.~T., \& {Scudder}, J.~M. 2013, \mnras, 433, L59

\bibitem[{Peng} {et~al.}(2012)]{2012ApJ...757....4P}
{Peng}, Y.-j., {Lilly}, S.~J., {Renzini}, A., \& {Carollo}, M. 2012, \apj, 757, 4

\bibitem[{Peng} {et~al.}(2010)]{2010ApJ...721..193P}
{Peng}, Y.-j., {Lilly}, S.~J., {Kova{\v{c}}}, K., {et~al.} 2010, \apj, 721, 193

\bibitem[{Rodighiero} {et~al.}(2011)]{2011ApJ...739L..40R}
{Rodighiero}, G., {Daddi}, E., {Baronchelli}, I., {et~al.} 2011, \apjl, 739, L40

\bibitem[{Sanders} \& {Mirabel}(1996)]{1996ARA&A..34..749S}
{Sanders}, D.~B., \& {Mirabel}, I.~F. 1996, \araa, 34, 749

\bibitem[{Sargent} \& {Scoville}(1991)]{1991ApJ...366L...1S}
{Sargent}, A., \& {Scoville}, N. 1991, \apjl, 366, L1

\bibitem[{Scoville} {et~al.}(1991)]{1991ApJ...366L...5S}
{Scoville}, N.~Z., {Sargent}, A.~I., {Sanders}, D.~B., \& {Soifer}, B.~T. 1991, \apjl, 366, L5

\bibitem[{Scudder} {et~al.}(2012)]{2012MNRAS.426..549S}
{Scudder}, J.~M., {Ellison}, S.~L., {Torrey}, P., {Patton}, D.~R., \& {Mendel}, J.~T. 2012, \mnras, 426, 549

\bibitem[{Steffen} {et~al.}(2021)]{2021ApJ...909..120S}
{Steffen}, J.~L., {Fu}, H., {Comerford}, J.~M., {et~al.} 2021, \apj, 909, 120

\bibitem[{Toomre} \& {Toomre}(1972)]{1972ApJ...178..623T}
{Toomre}, A., \& {Toomre}, J. 1972, \apj, 178, 623

\bibitem[{Xu} {et~al.}(2021)]{2021ApJ...918...55X}
{Xu}, C.~K., {Lisenfeld}, U., {Gao}, Y., \& {Renaud}, F. 2021, \apj, 918, 55

\bibitem[{Xu} {et~al.}(2010)]{2010ApJ...713..330X}
{Xu}, C.~K., {Domingue}, D., {Cheng}, Y.-W., {et~al.} 2010, \apj, 713, 330

\bibitem[{Xu} \& {Sulentic}(1991)]{1991ApJ...374..407X}
{Xu}, C., \& {Sulentic}, J.~W. 1991, \apj, 374, 407

\bibitem[{Yang} {et~al.}(2012)]{2012ApJ...752...41Y}
{Yang}, X., {Mo}, H.~J., {van den Bosch}, F.~C., {Zhang}, Y., \& {Han}, J. 2012, \apj, 752, 41

\bibitem[{Zuo} {et~al.}(2018)]{2018ApJS..237....2Z}
{Zuo}, P., {Xu}, C.~K., {Yun}, M.~S., {et~al.} 2018, \apjs, 237, 2

\end{thebibliography}



\appendix
\clearpage
\begin{table*}
\caption[]{Spiral galaxies in H-KPAIR\label{tab:1}}
\begin{tabular}{lrcccrrrrr}
\hline\hline
name & $z$ & pair\_type & bulge\_type & is\_SFG & $\rm r_{mag}$ & B/T & $\rm \log M_{star}$ & $\rm sSFR_{enh}$ & $\rm N_{1Mpc}$ \\
&  &  &  &  & (mag) &  & ($\rm M_{\odot}$) & (dex) & \\
(1) & (2) & (3) & (4) & (5) & (6) & (7) & (8) & (9) & (10)\\
\hline
J00202580$+$0049350 & 0.0149  & SE & p & 1 & 13.66  & 0.11  & 10.44  & 0.34  & 2 \\
J01183417$-$0013416 & 0.0453  & SS & p & 1 & 15.60  & 0.37  & 10.97  & 1.34  & 6 \\
J01183556$-$0013594 & 0.0455  & SS & c & 1 & 14.96  & 0.01  & 10.57  & 0.57  & 6 \\
J02110638$-$0039191 & 0.0177  & SS & c & 1 & 14.50  & 0.27  & 10.42  & 0.45  & 4 \\
J02110832$-$0039171 & 0.0181  & SS & c & 0 & 13.62  & 0.73  & 10.63  & $-$1.19  & 4 \\
J03381222$+$0110088 & 0.0392  & SE & c & 1 & 15.19  & 0.12  & 10.67  & 0.75  & 10 \\
J07543194$+$1648214 & 0.0459  & SS & p & 1 & 14.63  & 0.03  & 10.95  & 0.71  & 3 \\
J07543221$+$1648349 & 0.0462  & SS & p & 1 & 14.53  & 0.15  & 11.15  & 0.62  & 3 \\
J08083377$+$3854534 & 0.0402  & SE & c & 1 & 15.45  & 0.86  & 10.73  & 0.11  & 16 \\
J08233266$+$2120171 & 0.0181  & SS & p & 1 & 14.42  & 0.26  & 10.08  & 0.39  & 8 \\
J08233421$+$2120515 & 0.0181  & SS & c & 1 & 13.80  & 0.39  & 10.35  & 0.55  & 8 \\
J08291491$+$5531227 & 0.0251  & SS & p & 1 & 14.09  & 0.03  & 10.56  & 0.32  & 3 \\
J08292083$+$5531081 & 0.0252  & SS & p & 1 & 14.04  & 0.11  & 10.71  & 0.18  & 3 \\
J08364482$+$4722188 & 0.0526  & SS & p & 1 & 15.18  & 0.19  & 11.02  & $-$0.46  & 9 \\
J08364588$+$4722100 & 0.0526  & SS & c & 0 & 14.96  & 0.87  & 11.17  & $-$0.67  & 9 \\
J08381759$+$3054534 & 0.0476  & SS & c & 1 & 15.79  & 0.75  & 10.77  & 0.24  & 3 \\
J08381795$+$3055011 & 0.0481  & SS & c & 1 & 15.09  & 0.73  & 11.08  & $-$0.37  & 3 \\
J08390125$+$3613042 & 0.0548  & SE & c & 1 & 15.51  & 0.71  & 10.94  & $-$0.06  & 6 \\
J08414959$+$2642578 & 0.0848  & SE & p & 1 & 15.43  & 0.46  & 11.44  & $-$0.54  & 12 \\
J09060498$+$5144071 & 0.0291  & SE & c & 1 & 14.34  & 0.14  & 10.60  & 0.08  & 4 \\
J09123676$+$3547462 & 0.0235  & SE & c & 0 & 15.04  & 0.37  & 10.28  & $-$1.15  & 6 \\
J09134606$+$4742001 & 0.0527  & SE & c & 1 & 14.89  & 0.43  & 11.05  & 0.11  & 23 \\
J09155467$+$4419510 & 0.0396  & SS & p & 1 & 14.97  & 0.00  & 10.76  & 1.11  & 2 \\
J09155552$+$4419580 & 0.0396  & SS & p & 1 & 14.14  & 0.20  & 11.11  & 1.36  & 2 \\
J09264111$+$0447247 & 0.0891  & SS & c & 1 & 16.21  & 1.00  & 11.12  & $-$0.52  & 5 \\
J09264137$+$0447260 & 0.0907  & SS & c & 0 & 16.17  & 0.70  & 11.40  & $-$0.82  & 5 \\
J09374413$+$0245394 & 0.0242  & SE & c & 1 & 12.85  & 0.09  & 11.24  & 0.60  & 3 \\
J10100079$+$5440198 & 0.0460  & SS & c & 1 & 14.90  & 0.00  & 10.89  & 0.79  & 23 \\
J10100212$+$5440279 & 0.0463  & SS & c & 1 & 15.67  & 0.29  & 10.85  & 0.29  & 23 \\
J10155257$+$0657330 & 0.0299  & SE & p & 1 & 15.09  & 0.33  & 10.51  & $-$0.82  & 3 \\
J10205188$+$4831096 & 0.0530  & SE & c & 1 & 16.11  & 0.34  & 10.60  & 0.05  & 2 \\
J10225647$+$3446564 & 0.0554  & SS & p & 1 & 15.99  & 0.35  & 10.67  & 0.41  & 2 \\
J10225655$+$3446468 & 0.0564  & SS & c & 0 & 15.03  & 0.93  & 11.02  & $-$0.87  & 2 \\
J10233658$+$4220477 & 0.0456  & SS & p & 1 & 15.00  & 0.14  & 10.79  & 0.49  & 6 \\
J10233684$+$4221037 & 0.0454  & SS & p & 1 & 15.84  & 0.63  & 10.55  & 0.43  & 6 \\
J10272950$+$0114490 & 0.0236  & SE & p & 1 & 14.34  & 0.52  & 10.48  & 0.32  & 5 \\
J10325316$+$5306536 & 0.0640  & SE & c & 1 & 15.49  & 0.72  & 11.05  & $-$0.41  & 19 \\
J10332972$+$4404342 & 0.0523  & SS & p & 1 & 14.79  & 0.26  & 11.12  & 0.84  & 4 \\
J10333162$+$4404212 & 0.0521  & SS & p & 1 & 15.65  & 0.13  & 10.95  & 0.24  & 4 \\
J10364274$+$5447356 & 0.0458  & SE & c & 0 & 15.34  & 0.79  & 10.81  & $-$1.12  & 7 \\
J10392338$+$3904501 & 0.0435  & SE & p & 0 & 15.26  & 0.38  & 10.77  & $-$1.11  & 2 \\
J10435053$+$0645466 & 0.0287  & SS & c & 1 & 14.26  & 0.11  & 10.55  & 0.70  & 5 \\
J10435268$+$0645256 & 0.0281  & SS & p & 1 & 14.66  & 0.23  & 10.48  & $-$0.25  & 5 \\
J10452478$+$3910298 & 0.0268  & SE & c & 1 & 14.44  & 0.46  & 10.68  & $-$0.34  & 4 \\
J10514450$+$5101303 & 0.0238  & SE & c & 0 & 14.06  & 0.56  & 10.85  & $-$0.76  & 8 \\
J10595915$+$0857357 & 0.0627  & SE & c & 0 & 15.63  & 0.48  & 10.90  & $-$0.71  & 9 \\
continue...\\
\hline
\end{tabular}
\end{table*}

\begin{table*}
\caption*{Table 1 continue}
\begin{tabular}{lrcccrrrrr}
\hline\hline
name & $z$ & pair\_type & bulge\_type & is\_SFG & $\rm r_{mag}$ & B/T & $\rm \log M_{star}$ & $\rm sSFR_{enh}$ & $\rm N_{1Mpc}$ \\
&  &  &  &  & (mag) &  & ($\rm M_{\odot}$) & (dex) & \\
(1) & (2) & (3) & (4) & (5) & (6) & (7) & (8) & (9) & (10)\\
\hline
J11014364$+$5720336 & 0.0478  & SE & c & 1 & 15.85  & 0.55  & 10.58  & 0.14  & 6 \\
J11064944$+$4751119 & 0.0643  & SS & c & 0 & 15.48  & 0.87  & 11.02  & $-$0.83  & 4 \\
J11065068$+$4751090 & 0.0654  & SS & p & 1 & 15.23  & 0.13  & 11.09  & 0.40  & 4 \\
J11204657$+$0028142 & 0.0255  & SS & c & 1 & 14.37  & 0.96  & 10.54  & $-$0.12  & 7 \\
J11204801$+$0028068 & 0.0256  & SS & c & 0 & 13.86  & 0.47  & 10.91  & $-$0.91  & 7 \\
J11251704$+$0227007 & 0.0504  & SS & c & 0 & 15.73  & 0.52  & 10.79  & $-$0.73  & 5 \\
J11251716$+$0226488 & 0.0507  & SS & c & 1 & 15.01  & 0.59  & 11.04  & $-$0.16  & 5 \\
J11273289$+$3604168 & 0.0351  & SS & c & 1 & 14.36  & 0.54  & 10.90  & $-$0.13  & 5 \\
J11273467$+$3603470 & 0.0351  & SS & c & 1 & 13.86  & 0.11  & 11.23  & 0.39  & 5 \\
J11375801$+$4728143 & 0.0339  & SE & c & 0 & 14.64  & 0.59  & 10.81  & $-$0.70  & 5 \\
J11440433$+$3332339 & 0.0315  & SE & p & 1 & 15.38  & 0.15  & 10.40  & 0.24  & 16 \\
J11484370$+$3547002 & 0.0641  & SS & p & 1 & 16.50  & 0.23  & 10.97  & 0.38  & 5 \\
J11484525$+$3547092 & 0.0636  & SS & c & 1 & 14.77  & 0.13  & 11.27  & 0.53  & 5 \\
J11501333$+$3746107 & 0.0550  & SS & c & 1 & 15.56  & 0.54  & 10.88  & $-$0.32  & 2 \\
J11501399$+$3746306 & 0.0555  & SS & c & 1 & 15.19  & 0.10  & 10.95  & 0.09  & 2 \\
J11505764$+$1444200 & 0.0572  & SE & c & 0 & 15.94  & 0.59  & 10.92  & $-$0.96  & 5 \\
J11542299$+$4932509 & 0.0702  & SE & c & 0 & 15.92  & 0.21  & 10.98  & $-$0.80  & 3 \\
J12020424$+$5342317 & 0.0647  & SE & c & 1 & 16.01  & 0.89  & 10.93  & 0.06  & 3 \\
J12054066$+$0135365 & 0.0220  & SE & p & 0 & 14.52  & 0.26  & 10.33  & $-$1.09  & 16 \\
J12115507$+$4039182 & 0.0229  & SS & c & 1 & 14.51  & 0.40  & 10.44  & 0.08  & 5 \\
J12115648$+$4039184 & 0.0235  & SS & c & 1 & 14.87  & 0.13  & 10.42  & 0.60  & 5 \\
J12191866$+$1201054 & 0.0268  & SE & p & 1 & 15.14  & 0.37  & 10.20  & $-$0.01  & 7 \\
J12433887$+$4405399 & 0.0418  & SE & c & 1 & 14.84  & 0.15  & 10.89  & $-$0.13  & 3 \\
J12525011$+$4645272 & 0.0613  & SE & p & 1 & 15.47  & 0.10  & 11.08  & $-$0.30  & 4 \\
J13011662$+$4803366 & 0.0303  & SS & p & 1 & 14.55  & 0.50  & 10.49  & 0.64  & 6 \\
J13011835$+$4803304 & 0.0298  & SS & p & 1 & 15.16  & 0.22  & 10.31  & 0.65  & 6 \\
J13082737$+$0422125 & 0.0255  & SS & c & 1 & 15.61  & 0.03  & 9.88  & $-$0.30  & 1 \\
J13082964$+$0422045 & 0.0257  & SS & c & 1 & 14.80  & 0.40  & 10.29  & $-$0.38  & 1 \\
J13131470$+$3910382 & 0.0716  & SE & c & 1 & 16.36  & 0.43  & 10.95  & $-$0.48  & 15 \\
J13151386$+$4424264 & 0.0359  & SS & c & 1 & 14.55  & 0.20  & 10.74  & 0.00  & 3 \\
J13151726$+$4424255 & 0.0357  & SS & c & 1 & 14.26  & 0.94  & 11.08  & 0.70  & 3 \\
J13153076$+$6207447 & 0.0306  & SS & p & 1 & 14.59  & 0.26  & 10.59  & 0.64  & 2 \\
J13153506$+$6207287 & 0.0306  & SS & p & 1 & 14.57  & 0.16  & 10.77  & 1.59  & 2 \\
J13325525$-$0301347 & 0.0493  & SS & p & 1 & 15.61  & 0.51  & 10.67  & 0.62  & 3 \\
J13325655$-$0301395 & 0.0483  & SS & c & 1 & 14.81  & 0.05  & 10.93  & 0.26  & 3 \\
J13462001$-$0325407 & 0.0248  & SE & c & 0 & 13.97  & 0.68  & 10.79  & $-$0.62  & 3 \\
J14003661$-$0254327 & 0.0256  & SS & c & 0 & 14.26  & 0.70  & 10.72  & $-$1.33  & 12 \\
J14003796$-$0254227 & 0.0269  & SS & c & 0 & 14.76  & 0.49  & 10.66  & $-$0.82  & 13 \\
J14005783$+$4251203 & 0.0327  & SS & p & 1 & 15.10  & 0.02  & 10.76  & 0.82  & 2 \\
J14005879$+$4250427 & 0.0335  & SS & p & 1 & 15.25  & 0.38  & 10.65  & 0.95  & 2 \\
J14055079$+$6542598 & 0.0306  & SE & p & 1 & 15.49  & 0.29  & 10.34  & $-$0.06  & 7 \\
J14062157$+$5043303 & 0.0065  & SE & p & 1 & 12.00  & 0.05  & 10.18  & 0.17  & 3 \\
J14070703$-$0234513 & 0.0586  & SE & c & 1 & 15.55  & 0.13  & 11.05  & $-$0.19  & 4 \\
J14234238$+$3400324 & 0.0136  & SS & p & 1 & 13.55  & 0.28  & 10.11  & 0.29  & 4 \\
J14234632$+$3401012 & 0.0126  & SS & p & 1 & 13.91  & 0.17  & 10.26  & $-$0.36  & 4 \\
continue...\\
\hline
\end{tabular}
\end{table*}

\begin{table*}
\caption*{Table 1 continue}
\begin{tabular}{lrcccrrrrr}
\hline\hline
name & $z$ & pair\_type & bulge\_type & is\_SFG & $\rm r_{mag}$ & B/T & $\rm \log M_{star}$ & $\rm sSFR_{enh}$ & $\rm N_{1Mpc}$ \\
&  &  &  &  & (mag) &  & ($\rm M_{\odot}$) & (dex) & \\
(1) & (2) & (3) & (4) & (5) & (6) & (7) & (8) & (9) & (10)\\
\hline
J14245831$-$0303597 & 0.0525  & SS & c & 1 & 15.02  & 0.05  & 10.92  & 0.04  & 3 \\
J14245913$-$0304012 & 0.0535  & SS & c & 0 & 15.00  & 0.18  & 11.18  & $-$0.67  & 3 \\
J14250739$+$0313560 & 0.0375  & SE & c & 0 & 15.64  & 0.55  & 10.44  & $-$1.10  & 5 \\
J14294766$+$3534275 & 0.0290  & SS & c & 1 & 13.75  & 0.56  & 11.04  & $-$0.17  & 8 \\
J14295031$+$3534122 & 0.0296  & SS & p & 1 & 14.62  & 0.14  & 10.67  & $-$0.07  & 9 \\
J14334683$+$4004512 & 0.0260  & SS & c & 1 & 13.48  & 0.44  & 10.99  & 0.53  & 4 \\
J14334840$+$4005392 & 0.0264  & SS & c & 1 & 13.91  & 0.80  & 10.83  & 0.30  & 4 \\
J14442055$+$1207429 & 0.0304  & SS & c & 1 & 14.44  & 0.42  & 10.83  & 0.15  & 3 \\
J14442079$+$1207552 & 0.0314  & SS & c & 1 & 13.59  & 0.28  & 11.17  & 0.16  & 3 \\
J15002500$+$4317131 & 0.0316  & SE & c & 0 & 14.57  & 0.40  & 10.80  & $-$1.31  & 4 \\
J15053137$+$3427534 & 0.0745  & SE & c & 0 & 15.80  & 0.84  & 11.03  & $-$0.71  & 9 \\
J15064391$+$0346364 & 0.0363  & SS & c & 0 & 14.28  & 0.54  & 11.01  & $-$0.66  & 8 \\
J15064579$+$0346214 & 0.0352  & SS & c & 1 & 14.68  & 0.46  & 10.93  & 0.54  & 8 \\
J15101587$+$5810425 & 0.0303  & SS & c & 0 & 14.63  & 0.35  & 10.73  & $-$0.97  & 1 \\
J15101776$+$5810375 & 0.0317  & SS & p & 1 & 15.65  & 0.29  & 10.53  & 0.05  & 1 \\
J15144544$+$0403587 & 0.0386  & SS & c & 0 & 15.04  & 0.52  & 10.91  & $-$0.50  & 14 \\
J15144697$+$0403576 & 0.0392  & SS & c & 0 & 14.82  & 0.48  & 10.85  & $-$0.95  & 14 \\
J15233768$+$3749030 & 0.0234  & SE & p & 1 & 15.04  & 0.16  & 10.18  & $-$0.08  & 3 \\
J15264774$+$5915464 & 0.0447  & SE & c & 0 & 15.07  & 0.81  & 10.82  & $-$1.00  & 7 \\
J15281276$+$4255474 & 0.0188  & SS & p & 1 & 13.12  & 0.31  & 11.04  & $-$0.16  & 8 \\
J15281667$+$4256384 & 0.0180  & SS & c & 0 & 13.40  & 0.77  & 10.77  & $-$0.98  & 8 \\
J15523393$+$4620237 & 0.0610  & SE & c & 1 & 15.60  & 0.27  & 10.94  & 0.22  & 7 \\
J15562191$+$4757172 & 0.0191  & SE & p & 1 & 14.71  & 0.04  & 10.19  & 0.23  & 9 \\
J15583749$+$3227379 & 0.0494  & SS & p & 0 & 16.15  & 0.47  & 10.60  & $-$1.14  & 3 \\
J15583784$+$3227471 & 0.0485  & SS & p & 1 & 15.18  & 0.60  & 10.91  & 0.32  & 3 \\
J16024254$+$4111499 & 0.0335  & SS & c & 1 & 14.14  & 0.08  & 10.78  & 0.77  & 5 \\
J16024475$+$4111589 & 0.0333  & SS & p & 1 & 15.18  & 0.02  & 10.46  & 0.43  & 5 \\
J16080559$+$2529091 & 0.0415  & SS & p & 1 & 15.02  & 0.15  & 10.97  & 0.08  & 8 \\
J16080648$+$2529066 & 0.0423  & SS & p & 1 & 14.71  & 0.39  & 11.23  & $-$0.13  & 8 \\
J16082261$+$2328459 & 0.0409  & SS & c & 1 & 15.00  & 0.04  & 10.39  & 0.20  & 4 \\
J16082354$+$2328240 & 0.0408  & SS & c & 1 & 15.50  & 0.25  & 10.72  & 0.57  & 4 \\
J16145418$+$3711064 & 0.0582  & SE & c & 0 & 14.64  & 0.74  & 11.17  & $-$0.85  & 8 \\
J16282497$+$4110064 & 0.0330  & SS & c & 0 & 14.01  & 0.41  & 10.93  & $-$0.47  & 21 \\
J16282756$+$4109395 & 0.0318  & SS & c & 0 & 14.19  & 0.55  & 10.89  & $-$0.44  & 31 \\
J16354293$+$2630494 & 0.0701  & SE & c & 0 & 15.22  & 0.66  & 11.27  & $-$0.59  & 28 \\
J16372583$+$4650161 & 0.0578  & SS & c & 1 & 14.99  & 0.49  & 11.30  & $-$0.37  & 5 \\
J16372754$+$4650054 & 0.0568  & SS & p & 1 & 15.14  & 0.02  & 10.98  & $-$0.09  & 5 \\
J17020378$+$1859495 & 0.0558  & SE & c & 1 & 16.55  & 0.36  & 10.72  & $-$0.78  & 6 \\
J17045089$+$3448530 & 0.0572  & SS & c & 1 & 15.99  & 0.48  & 10.75  & 0.80  & 4 \\
J17045097$+$3449020 & 0.0563  & SS & p & 1 & 15.26  & 0.28  & 10.98  & 1.04  & 4 \\
J20471908$+$0019150 & 0.0130  & SE & c & 1 & 11.85  & 0.34  & 11.09  & $-$0.13  & 7 \\
\hline
\end{tabular}
\tablecomments{\hsize}{The columns are: (1) galaxy name; (2) redshift; (3) pair type:``S'' for Spiral+Spiral pair, ``S'' for Spiral+Elliptical pair; (4) if the galaxy is an SFG, 1 for true and 0 for false; (5) bulge type: ``p'' for pseudobulge, ``c'' for classical-bulge; (6) magnitude of $r$-band; (7) Bulge-to-Total ratio; (8) stellar mass; (9) specific star formation rate enhancement; (10) number of galaxies brighter than $\rm M_{r}=-19.7\; mag$ within a $\rm D=1\;Mpc$ circle centered on the galaxy.}
\end{table*}

\begin{table*}
\caption[]{Aperture photometry and model fits of the H-KPAIR sample\label{tab:2}}
\begin{tabular}{lcrrrrrrrrrr}
\hline\hline
name & type & ${\rm m}_u$ & ${\rm e}_u$ & ${\rm m}_r$ & ${\rm e}_r$ & ${\rm m}_i$ & ${\rm e}_i$ & ${\rm m}_{{\rm K}_s}$ & ${\rm e}_{{\rm K}_s}$ & Age & ${\rm A}_v$ \\ 
& & (mag) & (mag) & (mag) & (mag) & (mag) & (mag) & (mag) & (mag) & (Gyr) & (mag)\\
(1) & (2) & (3) & (4) & (5) & (6) & (7) & (8) & (9) & (10) & (11) & (12)\\
\hline
J00202580$+$0049350 & S & 16.86  & 0.02  & 14.51  & 0.01  & 14.09  & 0.01  & 13.34  & 0.02  & 0.56  & 1.25  \\
J00202748$+$0050009 & E & 16.48  & 0.02  & 13.87  & 0.01  & 13.44  & 0.01  & 12.71  & 0.01  & 3.21  & 0.61  \\
J01183417$-$0013416 & S & 18.51  & 0.03  & 16.17  & 0.01  & 15.65  & 0.01  & 14.18  & 0.07  & 0.05  & 2.85  \\
J01183556$-$0013594 & S & 18.17  & 0.03  & 16.55  & 0.01  & 16.24  & 0.01  & 15.66  & 0.05  & 0.09  & 1.54  \\
J02110638$-$0039191 & S & 17.75  & 0.03  & 15.25  & 0.01  & 14.77  & 0.01  & 13.50  & 0.02  & 0.10  & 2.77  \\
J02110832$-$0039171 & S & 16.30  & 0.02  & 13.91  & 0.01  & 13.54  & 0.01  & 12.92  & 0.01  & 1.61  & 0.67  \\
J03381222$+$0110088 & S & 18.67  & 0.03  & 16.39  & 0.01  & 15.92  & 0.01  & 14.97  & 0.04  & 0.17  & 2.02  \\
J03381299$+$0109414 & E & 18.94  & 0.03  & 16.09  & 0.01  & 15.63  & 0.01  & 14.76  & 0.08  & 6.31  & 0.61  \\
J07543194$+$1648214 & S & 18.17  & 0.03  & 15.93  & 0.01  & 15.52  & 0.01  & 14.44  & 0.28  & 0.09  & 2.42  \\
J07543221$+$1648349 & S & 17.75  & 0.02  & 15.64  & 0.01  & 15.23  & 0.01  & 14.35  & 0.26  & 0.13  & 2.01  \\
J08083377$+$3854534 & S & 18.34  & 0.03  & 16.01  & 0.01  & 15.61  & 0.01  & 14.84  & 0.04  & 0.46  & 1.34  \\
J08083563$+$3854522 & E & 18.27  & 0.03  & 15.49  & 0.01  & 15.05  & 0.01  & 14.24  & 0.02  & 6.07  & 0.53  \\
J08233266$+$2120171 & S & 16.10  & 0.02  & 14.73  & 0.01  & 14.53  & 0.01  & 14.07  & 0.09  & 0.07  & 1.26  \\
J08233421$+$2120515 & S & 16.07  & 0.02  & 14.47  & 0.01  & 14.21  & 0.01  & 13.67  & 0.06  & 0.09  & 1.48  \\
J08291491$+$5531227 & S & 17.57  & 0.03  & 15.37  & 0.01  & 14.97  & 0.01  & 14.25  & 0.02  & 0.34  & 1.36  \\
J08292083$+$5531081 & S & 17.24  & 0.02  & 14.84  & 0.01  & 14.44  & 0.01  & 13.70  & 0.01  & 0.72  & 1.16  \\
J08364482$+$4722188 & S & 18.99  & 0.03  & 16.28  & 0.01  & 15.83  & 0.01  & 14.95  & 0.03  & 1.74  & 1.11  \\
J08364588$+$4722100 & S & 18.66  & 0.03  & 15.84  & 0.01  & 15.41  & 0.01  & 14.57  & 0.02  & 7.35  & 0.48  \\
J08381759$+$3054534 & S & 18.73  & 0.03  & 16.34  & 0.01  & 15.85  & 0.01  & 14.84  & 0.02  & 0.19  & 2.09  \\
J08381795$+$3055011 & S & 19.00  & 0.03  & 16.07  & 0.01  & 15.52  & 0.01  & 14.31  & 0.01  & 0.61  & 2.05  \\
J08385973$+$3613164 & E & 18.58  & 0.03  & 15.74  & 0.01  & 15.30  & 0.01  & 14.53  & 0.03  & 13.00  & 0.12  \\
J08390125$+$3613042 & S & 19.00  & 0.03  & 16.35  & 0.01  & 15.89  & 0.01  & 15.01  & 0.05  & 0.98  & 1.32  \\
J08414959$+$2642578 & S & 19.65  & 0.03  & 16.69  & 0.01  & 16.16  & 0.01  & 15.21  & 0.03  & 6.62  & 0.75  \\
J08415054$+$2642475 & E & 20.02  & 0.04  & 17.09  & 0.01  & 16.63  & 0.01  & 16.08  & 0.05  & 13.00  & 0.00  \\
J09060283$+$5144411 & E & 17.61  & 0.02  & 14.91  & 0.01  & 14.52  & 0.01  & 13.74  & 0.02  & 4.52  & 0.56  \\
J09060498$+$5144071 & S & 17.99  & 0.03  & 15.59  & 0.01  & 15.20  & 0.01  & 14.49  & 0.03  & 0.86  & 1.05  \\
J09123636$+$3547180 & E & 17.15  & 0.02  & 14.58  & 0.01  & 14.18  & 0.01  & 13.56  & 0.01  & 7.19  & 0.12  \\
J09123676$+$3547462 & S & 17.65  & 0.02  & 15.25  & 0.01  & 14.87  & 0.01  & 14.34  & 0.02  & 4.52  & 0.13  \\
J09134461$+$4742165 & E & 18.57  & 0.03  & 15.75  & 0.01  & 15.31  & 0.01  & 14.56  & 0.02  & 13.00  & 0.02  \\
J09134606$+$4742001 & S & 17.54  & 0.02  & 15.98  & 0.01  & 15.49  & 0.01  & 14.43  & 0.01  & 0.04  & 1.86  \\
J09155467$+$4419510 & S & 18.19  & 0.03  & 15.85  & 0.01  & 15.40  & 0.01  & 14.22  & 0.01  & 0.09  & 2.60  \\
J09155552$+$4419580 & S & 17.58  & 0.02  & 15.45  & 0.01  & 15.03  & 0.01  & 13.93  & 0.01  & 0.07  & 2.39  \\
J09264111$+$0447247 & S & 20.02  & 0.04  & 17.18  & 0.01  & 16.73  & 0.01  & 15.98  & 0.15  & 13.00  & 0.00  \\
J09264137$+$0447260 & S & 20.03  & 0.04  & 17.11  & 0.01  & 16.65  & 0.01  & 15.69  & 0.11  & 4.26  & 0.93  \\
J09374413$+$0245394 & S & 16.77  & 0.02  & 14.41  & 0.01  & 13.97  & 0.01  & 12.83  & 0.01  & 0.10  & 2.57  \\
J09374506$+$0244504 & E & 16.79  & 0.02  & 14.00  & 0.01  & 13.57  & 0.01  & 12.72  & 0.01  & 4.36  & 0.72  \\
J10100079$+$5440198 & S & 18.61  & 0.03  & 16.46  & 0.01  & 16.01  & 0.01  & 14.91  & 0.01  & 0.07  & 2.39  \\
J10100212$+$5440279 & S & 18.46  & 0.03  & 16.26  & 0.01  & 15.83  & 0.01  & 14.96  & 0.01  & 0.18  & 1.86  \\
J10155257$+$0657330 & S & 17.87  & 0.02  & 15.57  & 0.01  & 15.20  & 0.01  & 14.60  & 0.15  & 0.92  & 0.84  \\
J10155338$+$0657495 & E & 17.41  & 0.02  & 14.79  & 0.01  & 14.38  & 0.01  & 13.66  & 0.07  & 3.77  & 0.54  \\
J10205188$+$4831096 & S & 19.31  & 0.03  & 17.00  & 0.01  & 16.56  & 0.01  & 15.76  & 0.06  & 0.36  & 1.48  \\
J10205369$+$4831246 & E & 19.07  & 0.03  & 16.18  & 0.01  & 15.72  & 0.01  & 14.86  & 0.03  & 9.75  & 0.43  \\
J10225647$+$3446564 & S & 18.90  & 0.03  & 16.70  & 0.01  & 16.27  & 0.01  & 15.41  & 0.05  & 0.18  & 1.84  \\
J10225655$+$3446468 & S & 18.33  & 0.03  & 15.79  & 0.01  & 15.41  & 0.01  & 14.76  & 0.03  & 4.26  & 0.37  \\
J10233658$+$4220477 & S & 17.76  & 0.02  & 16.02  & 0.01  & 15.62  & 0.01  & 14.82  & 0.01  & 0.06  & 1.83  \\
J10233684$+$4221037 & S & 17.89  & 0.02  & 16.28  & 0.01  & 15.88  & 0.01  & 15.15  & 0.02  & 0.06  & 1.68  \\
continue...\\
\hline
\end{tabular}
\end{table*}

\begin{table*}
\caption*{Table2 continue}
\begin{tabular}{lcrrrrrrrrrr}
\hline\hline
name & type & ${\rm m}_u$ & ${\rm e}_u$ & ${\rm m}_r$ & ${\rm e}_r$ & ${\rm m}_i$ & ${\rm e}_i$ & ${\rm m}_{{\rm K}_s}$ & ${\rm e}_{{\rm K}_s}$ & Age & ${\rm A}_v$ \\ 
& & (mag) & (mag) & (mag) & (mag) & (mag) & (mag) & (mag) & (mag) & (Gyr) & (mag)\\
(1) & (2) & (3) & (4) & (5) & (6) & (7) & (8) & (9) & (10) & (11) & (12)\\
\hline
J10272950$+$0114490 & S & 17.24  & 0.02  & 15.15  & 0.01  & 14.79  & 0.01  & 13.83  & 0.05  & 0.09  & 2.21  \\
J10272970$+$0115170 & E & 16.93  & 0.02  & 14.31  & 0.01  & 13.90  & 0.01  & 13.23  & 0.03  & 6.78  & 0.22  \\
J10325316$+$5306536 & S & 19.31  & 0.03  & 16.58  & 0.01  & 16.11  & 0.01  & 15.27  & 0.05  & 3.12  & 0.81  \\
J10325321$+$5306477 & E & 18.82  & 0.03  & 16.01  & 0.01  & 15.57  & 0.01  & 14.83  & 0.03  & 13.00  & 0.00  \\
J10332972$+$4404342 & S & 17.72  & 0.02  & 15.64  & 0.01  & 15.22  & 0.01  & 14.35  & 0.01  & 0.11  & 2.04  \\
J10333162$+$4404212 & S & 19.40  & 0.04  & 16.76  & 0.01  & 16.26  & 0.01  & 15.01  & 0.02  & 0.17  & 2.56  \\
J10364274$+$5447356 & S & 18.95  & 0.03  & 16.27  & 0.01  & 15.87  & 0.01  & 15.09  & 0.12  & 3.48  & 0.68  \\
J10364400$+$5447489 & E & 17.93  & 0.02  & 15.28  & 0.01  & 14.88  & 0.01  & 14.15  & 0.05  & 4.85  & 0.45  \\
J10392338$+$3904501 & S & 18.69  & 0.03  & 15.94  & 0.01  & 15.54  & 0.01  & 14.78  & 0.02  & 8.30  & 0.30  \\
J10392515$+$3904573 & E & 18.36  & 0.03  & 15.64  & 0.01  & 15.24  & 0.01  & 14.53  & 0.01  & 10.62  & 0.14  \\
J10435053$+$0645466 & S & 16.29  & 0.02  & 15.05  & 0.01  & 14.83  & 0.01  & 14.15  & 0.02  & 0.04  & 1.31  \\
J10435268$+$0645256 & S & 18.08  & 0.03  & 15.64  & 0.01  & 15.24  & 0.01  & 14.37  & 0.02  & 0.43  & 1.54  \\
J10452478$+$3910298 & S & 17.51  & 0.02  & 15.00  & 0.01  & 14.60  & 0.01  & 13.89  & 0.02  & 1.70  & 0.81  \\
J10452496$+$3909499 & E & 17.43  & 0.02  & 14.81  & 0.01  & 14.41  & 0.01  & 13.73  & 0.26  & 5.86  & 0.30  \\
J10514368$+$5101195 & E & 16.65  & 0.02  & 13.96  & 0.01  & 13.55  & 0.01  & 12.78  & 0.01  & 4.72  & 0.53  \\
J10514450$+$5101303 & S & 17.14  & 0.02  & 14.55  & 0.01  & 14.15  & 0.01  & 13.51  & 0.02  & 7.81  & 0.11  \\
J10595869$+$0857215 & E & 18.92  & 0.03  & 16.06  & 0.01  & 15.64  & 0.01  & 14.95  & 0.02  & 13.00  & 0.00  \\
J10595915$+$0857357 & S & 19.29  & 0.03  & 16.47  & 0.01  & 16.04  & 0.01  & 15.50  & 0.03  & 13.00  & 0.00  \\
J11014357$+$5720058 & E & 18.21  & 0.03  & 15.65  & 0.01  & 15.30  & 0.01  & 14.66  & 0.02  & 5.65  & 0.25  \\
J11014364$+$5720336 & S & 19.02  & 0.03  & 16.35  & 0.01  & 15.97  & 0.01  & 15.25  & 0.03  & 5.89  & 0.38  \\
J11064944$+$4751119 & S & 18.50  & 0.03  & 15.89  & 0.01  & 15.55  & 0.01  & 14.85  & 0.01  & 4.69  & 0.41  \\
J11065068$+$4751090 & S & 18.88  & 0.03  & 16.54  & 0.01  & 16.10  & 0.01  & 15.20  & 0.02  & 0.24  & 1.80  \\
J11204657$+$0028142 & S & 16.73  & 0.02  & 14.60  & 0.01  & 14.31  & 0.01  & 13.71  & 0.02  & 0.48  & 1.03  \\
J11204801$+$0028068 & S & 17.23  & 0.02  & 14.47  & 0.01  & 14.04  & 0.01  & 13.25  & 0.01  & 6.15  & 0.49  \\
J11251704$+$0227007 & S & 18.82  & 0.03  & 16.14  & 0.01  & 15.73  & 0.01  & 14.99  & 0.03  & 5.13  & 0.46  \\
J11251716$+$0226488 & S & 18.63  & 0.03  & 15.85  & 0.01  & 15.41  & 0.01  & 14.63  & 0.02  & 7.92  & 0.37  \\
J11273289$+$3604168 & S & 17.68  & 0.02  & 15.12  & 0.01  & 14.74  & 0.01  & 13.97  & 0.02  & 1.36  & 1.01  \\
J11273467$+$3603470 & S & 17.50  & 0.02  & 15.01  & 0.01  & 14.57  & 0.01  & 13.59  & 0.01  & 0.32  & 1.82  \\
J11375476$+$4727588 & E & 17.35  & 0.02  & 14.69  & 0.01  & 14.29  & 0.01  & 13.53  & 0.01  & 4.00  & 0.58  \\
J11375801$+$4728143 & S & 17.79  & 0.02  & 15.09  & 0.01  & 14.68  & 0.01  & 13.93  & 0.01  & 5.65  & 0.44  \\
J11440335$+$3332062 & E & 17.89  & 0.03  & 15.18  & 0.01  & 14.76  & 0.01  & 13.84  & 0.02  & 1.02  & 1.38  \\
J11440433$+$3332339 & S & 17.83  & 0.03  & 15.77  & 0.01  & 15.40  & 0.01  & 14.65  & 0.04  & 0.19  & 1.62  \\
J11484370$+$3547002 & S & 20.53  & 0.06  & 17.90  & 0.01  & 17.25  & 0.01  & 15.59  & 0.04  & 0.05  & 3.26  \\
J11484525$+$3547092 & S & 18.93  & 0.03  & 16.48  & 0.01  & 16.03  & 0.01  & 15.20  & 0.03  & 0.57  & 1.38  \\
J11501333$+$3746107 & S & 18.89  & 0.03  & 16.22  & 0.01  & 15.80  & 0.01  & 15.09  & 0.03  & 6.71  & 0.31  \\
J11501399$+$3746306 & S & 18.06  & 0.02  & 15.99  & 0.01  & 15.58  & 0.01  & 14.87  & 0.02  & 0.23  & 1.47  \\
J11505764$+$1444200 & S & 19.49  & 0.03  & 16.68  & 0.01  & 16.24  & 0.01  & 15.43  & 0.03  & 7.53  & 0.45  \\
J11505844$+$1444124 & E & 18.05  & 0.02  & 15.41  & 0.01  & 15.03  & 0.01  & 14.33  & 0.01  & 6.17  & 0.31  \\
J11542299$+$4932509 & S & 19.98  & 0.04  & 17.30  & 0.01  & 16.88  & 0.01  & 16.19  & 0.07  & 9.27  & 0.15  \\
J11542307$+$4932456 & E & 19.31  & 0.03  & 16.35  & 0.01  & 15.90  & 0.01  & 15.15  & 0.03  & 13.00  & 0.00  \\
J12020424$+$5342317 & S & 20.00  & 0.05  & 17.04  & 0.01  & 16.50  & 0.01  & 15.58  & 0.02  & 9.36  & 0.55  \\
J12020537$+$5342487 & E & 19.06  & 0.03  & 16.20  & 0.01  & 15.77  & 0.01  & 15.02  & 0.01  & 13.00  & 0.00  \\
J12054066$+$0135365 & S & 17.37  & 0.02  & 14.90  & 0.01  & 14.52  & 0.01  & 13.94  & 0.02  & 4.29  & 0.26  \\
J12054073$+$0134302 & E & 17.15  & 0.02  & 14.49  & 0.01  & 14.09  & 0.01  & 13.40  & 0.02  & 8.31  & 0.18  \\
J12115507$+$4039182 & S & 16.51  & 0.02  & 14.68  & 0.01  & 14.37  & 0.01  & 13.79  & 0.01  & 0.16  & 1.38  \\
J12115648$+$4039184 & S & 17.50  & 0.02  & 15.10  & 0.01  & 14.74  & 0.01  & 14.00  & 0.02  & 0.72  & 1.15  \\
continue...\\
\hline
\end{tabular}
\end{table*}

\begin{table*}
\caption*{Table2 continue}
\begin{tabular}{lcrrrrrrrrrr}
\hline\hline
name & type & ${\rm m}_u$ & ${\rm e}_u$ & ${\rm m}_r$ & ${\rm e}_r$ & ${\rm m}_i$ & ${\rm e}_i$ & ${\rm m}_{{\rm K}_s}$ & ${\rm e}_{{\rm K}_s}$ & Age & ${\rm A}_v$ \\ 
& & (mag) & (mag) & (mag) & (mag) & (mag) & (mag) & (mag) & (mag) & (Gyr) & (mag)\\
(1) & (2) & (3) & (4) & (5) & (6) & (7) & (8) & (9) & (10) & (11) & (12)\\
\hline
J12191719$+$1200582 & E & 17.82  & 0.03  & 15.17  & 0.01  & 14.77  & 0.01  & 14.09  & 0.03  & 7.98  & 0.18  \\
J12191866$+$1201054 & S & 17.63  & 0.02  & 15.69  & 0.01  & 15.33  & 0.01  & 14.66  & 0.04  & 0.17  & 1.52  \\
J12433887$+$4405399 & S & 18.73  & 0.03  & 16.07  & 0.01  & 15.66  & 0.01  & 14.95  & 0.03  & 6.43  & 0.32  \\
J12433936$+$4406046 & E & 18.00  & 0.02  & 15.31  & 0.01  & 14.92  & 0.01  & 14.19  & 0.02  & 6.70  & 0.33  \\
J12525011$+$4645272 & S & 18.98  & 0.03  & 16.48  & 0.01  & 16.05  & 0.01  & 15.37  & 0.03  & 2.10  & 0.69  \\
J12525212$+$4645294 & E & 18.92  & 0.03  & 16.11  & 0.01  & 15.63  & 0.01  & 14.83  & 0.02  & 8.83  & 0.36  \\
J13011662$+$4803366 & S & 16.20  & 0.02  & 15.00  & 0.01  & 14.80  & 0.01  & 14.29  & 0.02  & 0.05  & 1.18  \\
J13011835$+$4803304 & S & 17.04  & 0.02  & 15.56  & 0.01  & 15.31  & 0.01  & 14.78  & 0.03  & 0.07  & 1.40  \\
J13082737$+$0422125 & S & 17.67  & 0.02  & 15.98  & 0.01  & 15.70  & 0.01  & 15.36  & 0.06  & 0.28  & 0.75  \\
J13082964$+$0422045 & S & 17.36  & 0.02  & 15.32  & 0.01  & 14.98  & 0.01  & 14.41  & 0.03  & 0.37  & 1.07  \\
J13131429$+$3910360 & E & 19.15  & 0.03  & 16.26  & 0.01  & 15.84  & 0.01  & 15.13  & 0.02  & 13.00  & 0.00  \\
J13131470$+$3910382 & S & 19.54  & 0.03  & 16.78  & 0.01  & 16.40  & 0.01  & 15.75  & 0.03  & 13.00  & 0.00  \\
J13151386$+$4424264 & S & 17.40  & 0.02  & 15.46  & 0.01  & 15.12  & 0.01  & 14.38  & 0.19  & 0.13  & 1.74  \\
J13151726$+$4424255 & S & 17.41  & 0.02  & 15.02  & 0.01  & 14.59  & 0.01  & 13.43  & 0.08  & 0.11  & 2.57  \\
J13153076$+$6207447 & S & 17.33  & 0.02  & 15.56  & 0.01  & 15.23  & 0.01  & 14.06  & 0.02  & 0.04  & 2.13  \\
J13153506$+$6207287 & S & 16.62  & 0.02  & 15.13  & 0.01  & 14.81  & 0.01  & 13.49  & 0.01  & 0.02  & 2.00  \\
J13325525$-$0301347 & S & 18.60  & 0.03  & 16.37  & 0.01  & 15.95  & 0.01  & 15.05  & 0.02  & 0.17  & 1.93  \\
J13325655$-$0301395 & S & 18.27  & 0.03  & 16.19  & 0.01  & 15.79  & 0.01  & 15.12  & 0.02  & 0.27  & 1.34  \\
J13462001$-$0325407 & S & 17.21  & 0.02  & 14.62  & 0.01  & 14.22  & 0.01  & 13.48  & 0.02  & 2.57  & 0.71  \\
J13462215$-$0325057 & E & 17.50  & 0.02  & 14.82  & 0.01  & 14.38  & 0.01  & 13.60  & 0.01  & 3.34  & 0.70  \\
J14003661$-$0254327 & S & 17.54  & 0.02  & 14.91  & 0.01  & 14.50  & 0.01  & 13.79  & 0.01  & 4.94  & 0.41  \\
J14003796$-$0254227 & S & 17.70  & 0.02  & 15.13  & 0.01  & 14.73  & 0.01  & 13.98  & 0.02  & 1.85  & 0.86  \\
J14005783$+$4251203 & S & 18.53  & 0.03  & 15.94  & 0.01  & 15.43  & 0.01  & 14.20  & 0.03  & 0.16  & 2.54  \\
J14005879$+$4250427 & S & 17.95  & 0.02  & 15.78  & 0.01  & 15.32  & 0.01  & 14.16  & 0.03  & 0.06  & 2.47  \\
J14055079$+$6542598 & S & 18.47  & 0.03  & 15.98  & 0.01  & 15.54  & 0.01  & 14.72  & 0.09  & 0.66  & 1.34  \\
J14055334$+$6542277 & E & 17.58  & 0.02  & 15.07  & 0.01  & 14.70  & 0.01  & 14.14  & 0.06  & 7.91  & 0.00  \\
J14062157$+$5043303 & S & 14.02  & 0.02  & 12.46  & 0.01  & 12.20  & 0.01  & 11.78  & 0.01  & 0.13  & 1.19  \\
J14064127$+$5043239 & E & 15.08  & 0.02  & 12.58  & 0.01  & 12.16  & 0.01  & 11.59  & 0.01  & 7.61  & 0.00  \\
J14070703$-$0234513 & S & 19.56  & 0.04  & 16.73  & 0.01  & 16.24  & 0.01  & 15.39  & 0.04  & 5.96  & 0.61  \\
J14070720$-$0234402 & E & 19.35  & 0.03  & 16.52  & 0.01  & 16.08  & 0.01  & 15.26  & 0.03  & 9.04  & 0.39  \\
J14234238$+$3400324 & S & 15.37  & 0.02  & 13.97  & 0.01  & 13.77  & 0.01  & 13.36  & 0.02  & 0.09  & 1.22  \\
J14234632$+$3401012 & S & 16.35  & 0.02  & 14.18  & 0.01  & 13.78  & 0.01  & 13.01  & 0.02  & 0.24  & 1.56  \\
J14245831$-$0303597 & S & 17.81  & 0.02  & 15.85  & 0.01  & 15.50  & 0.01  & 14.90  & 0.44  & 0.24  & 1.26  \\
J14245913$-$0304012 & S & 19.41  & 0.03  & 16.38  & 0.01  & 15.87  & 0.01  & 14.92  & 0.43  & 12.91  & 0.48  \\
J14250552$+$0313590 & E & 17.84  & 0.02  & 15.28  & 0.01  & 14.90  & 0.01  & 14.07  & 0.01  & 0.88  & 1.27  \\
J14250739$+$0313560 & S & 18.70  & 0.03  & 16.31  & 0.01  & 15.96  & 0.01  & 15.10  & 0.03  & 0.38  & 1.57  \\
J14294766$+$3534275 & S & 17.10  & 0.02  & 14.42  & 0.01  & 14.00  & 0.01  & 13.25  & 0.02  & 5.03  & 0.49  \\
J14295031$+$3534122 & S & 17.73  & 0.02  & 15.27  & 0.01  & 14.88  & 0.01  & 14.20  & 0.03  & 1.55  & 0.79  \\
J14334683$+$4004512 & S & 17.29  & 0.02  & 14.71  & 0.01  & 14.27  & 0.01  & 13.37  & 0.01  & 0.69  & 1.45  \\
J14334840$+$4005392 & S & 16.33  & 0.02  & 14.45  & 0.01  & 14.11  & 0.01  & 13.21  & 0.01  & 0.07  & 2.02  \\
J14442055$+$1207429 & S & 17.28  & 0.02  & 15.11  & 0.01  & 14.74  & 0.01  & 13.96  & 0.02  & 0.22  & 1.62  \\
J14442079$+$1207552 & S & 17.59  & 0.02  & 14.86  & 0.01  & 14.44  & 0.01  & 13.66  & 0.01  & 5.77  & 0.47  \\
J15002374$+$4316559 & E & 17.60  & 0.02  & 14.88  & 0.01  & 14.45  & 0.01  & 13.62  & 0.01  & 3.06  & 0.81  \\
J15002500$+$4317131 & S & 17.81  & 0.02  & 15.05  & 0.01  & 14.63  & 0.01  & 13.83  & 0.02  & 6.08  & 0.50  \\
J15053137$+$3427534 & S & 18.98  & 0.03  & 16.36  & 0.01  & 16.01  & 0.01  & 15.47  & 0.02  & 13.00  & 0.00  \\
J15053183$+$3427526 & E & 19.10  & 0.03  & 16.24  & 0.01  & 15.83  & 0.01  & 15.12  & 0.01  & 13.00  & 0.00  \\
continue...\\
\hline
\end{tabular}
\end{table*}

\begin{table*}
\caption*{Table2 continue}
\begin{tabular}{lcrrrrrrrrrr}
\hline\hline
name & type & ${\rm m}_u$ & ${\rm e}_u$ & ${\rm m}_r$ & ${\rm e}_r$ & ${\rm m}_i$ & ${\rm e}_i$ & ${\rm m}_{{\rm K}_s}$ & ${\rm e}_{{\rm K}_s}$ & Age & ${\rm A}_v$ \\ 
& & (mag) & (mag) & (mag) & (mag) & (mag) & (mag) & (mag) & (mag) & (Gyr) & (mag)\\
(1) & (2) & (3) & (4) & (5) & (6) & (7) & (8) & (9) & (10) & (11) & (12)\\
\hline
J15064391$+$0346364 & S & 17.56  & 0.02  & 14.90  & 0.01  & 14.50  & 0.01  & 13.71  & 0.01  & 2.88  & 0.76  \\
J15064579$+$0346214 & S & 18.22  & 0.03  & 15.60  & 0.01  & 15.09  & 0.01  & 13.83  & 0.02  & 0.15  & 2.63  \\
J15101587$+$5810425 & S & 17.59  & 0.02  & 15.08  & 0.01  & 14.70  & 0.01  & 14.03  & 0.10  & 2.42  & 0.63  \\
J15101776$+$5810375 & S & 19.02  & 0.03  & 16.48  & 0.01  & 15.95  & 0.01  & 14.49  & 0.15  & 0.06  & 3.01  \\
J15144544$+$0403587 & S & 18.35  & 0.03  & 15.49  & 0.01  & 15.03  & 0.01  & 14.17  & 0.01  & 7.48  & 0.53  \\
J15144697$+$0403576 & S & 18.26  & 0.03  & 15.52  & 0.01  & 15.12  & 0.01  & 14.32  & 0.01  & 4.77  & 0.59  \\
J15233768$+$3749030 & S & 17.57  & 0.02  & 15.37  & 0.01  & 15.04  & 0.01  & 14.51  & 0.06  & 0.82  & 0.77  \\
J15233899$+$3748254 & E & 17.83  & 0.03  & 15.32  & 0.01  & 14.93  & 0.01  & 14.38  & 0.06  & 8.72  & 0.00  \\
J15264774$+$5915464 & S & 18.55  & 0.03  & 15.74  & 0.01  & 15.33  & 0.01  & 14.58  & 0.03  & 13.00  & 0.04  \\
J15264892$+$5915478 & E & 18.39  & 0.03  & 15.65  & 0.01  & 15.25  & 0.01  & 14.56  & 0.03  & 13.00  & 0.00  \\
J15281276$+$4255474 & S & 16.40  & 0.02  & 13.89  & 0.01  & 13.42  & 0.01  & 12.30  & 0.01  & 0.18  & 2.29  \\
J15281667$+$4256384 & S & 16.49  & 0.02  & 13.79  & 0.01  & 13.35  & 0.01  & 12.54  & 0.02  & 3.04  & 0.78  \\
J15523258$+$4620180 & E & 18.97  & 0.03  & 16.05  & 0.01  & 15.63  & 0.01  & 14.91  & 0.02  & 13.00  & 0.00  \\
J15523393$+$4620237 & S & 19.09  & 0.03  & 16.70  & 0.01  & 16.25  & 0.01  & 15.48  & 0.03  & 0.58  & 1.27  \\
J15562191$+$4757172 & S & 17.47  & 0.02  & 15.30  & 0.01  & 14.93  & 0.01  & 14.21  & 0.07  & 0.31  & 1.37  \\
J15562738$+$4757302 & E & 17.59  & 0.02  & 14.99  & 0.01  & 14.60  & 0.01  & 13.98  & 0.03  & 9.50  & 0.02  \\
J15583749$+$3227379 & S & 19.11  & 0.03  & 16.48  & 0.01  & 16.05  & 0.01  & 15.23  & 0.03  & 1.49  & 1.07  \\
J15583784$+$3227471 & S & 18.13  & 0.02  & 16.03  & 0.01  & 15.60  & 0.01  & 14.71  & 0.28  & 0.12  & 2.06  \\
J16024254$+$4111499 & S & 16.97  & 0.02  & 15.38  & 0.01  & 15.09  & 0.01  & 14.34  & 0.10  & 0.05  & 1.69  \\
J16024475$+$4111589 & S & 17.37  & 0.02  & 15.80  & 0.01  & 15.52  & 0.01  & 14.88  & 0.16  & 0.07  & 1.57  \\
J16080559$+$2529091 & S & 18.95  & 0.03  & 15.91  & 0.01  & 15.37  & 0.01  & 14.24  & 0.02  & 2.26  & 1.45  \\
J16080648$+$2529066 & S & 19.22  & 0.03  & 16.18  & 0.01  & 15.53  & 0.01  & 13.99  & 0.01  & 0.19  & 3.02  \\
J16082261$+$2328459 & S & 19.14  & 0.03  & 16.92  & 0.01  & 16.53  & 0.01  & 15.84  & 0.05  & 0.43  & 1.23  \\
J16082354$+$2328240 & S & 18.50  & 0.03  & 16.08  & 0.01  & 15.66  & 0.01  & 14.63  & 0.02  & 0.20  & 2.09  \\
J16145418$+$3711064 & S & 18.74  & 0.03  & 15.85  & 0.01  & 15.41  & 0.01  & 14.53  & 0.02  & 8.17  & 0.53  \\
J16145421$+$3711136 & E & 19.03  & 0.03  & 16.17  & 0.01  & 15.74  & 0.01  & 14.94  & 0.02  & 13.00  & 0.12  \\
J16282497$+$4110064 & S & 17.88  & 0.02  & 15.25  & 0.01  & 14.85  & 0.01  & 14.23  & 0.02  & 12.99  & 0.00  \\
J16282756$+$4109395 & S & 17.65  & 0.02  & 14.91  & 0.01  & 14.48  & 0.01  & 13.75  & 0.02  & 9.85  & 0.20  \\
J16354293$+$2630494 & S & 19.36  & 0.03  & 16.44  & 0.01  & 15.97  & 0.01  & 15.18  & 0.04  & 13.00  & 0.00  \\
J16354366$+$2630505 & E & 19.40  & 0.03  & 16.43  & 0.01  & 15.94  & 0.01  & 14.99  & 0.03  & 7.35  & 0.70  \\
J16372583$+$4650161 & S & 19.11  & 0.03  & 16.19  & 0.01  & 15.72  & 0.01  & 14.70  & 0.04  & 2.52  & 1.21  \\
J16372754$+$4650054 & S & 19.65  & 0.04  & 17.09  & 0.01  & 16.65  & 0.01  & 15.93  & 0.06  & 2.64  & 0.66  \\
J17020320$+$1900006 & E & 19.30  & 0.03  & 16.17  & 0.01  & 15.70  & 0.01  & 14.94  & 0.02  & 13.00  & 0.00  \\
J17020378$+$1859495 & S & 20.27  & 0.05  & 17.57  & 0.01  & 17.13  & 0.01  & 16.30  & 0.07  & 2.54  & 0.88  \\
J17045089$+$3448530 & S & 18.90  & 0.03  & 16.73  & 0.01  & 16.26  & 0.01  & 15.30  & 0.10  & 0.11  & 2.18  \\
J17045097$+$3449020 & S & 17.65  & 0.02  & 16.00  & 0.01  & 15.55  & 0.01  & 14.68  & 0.06  & 0.05  & 1.83  \\
J20471908$+$0019150 & S & 15.73  & 0.02  & 12.82  & 0.01  & 12.34  & 0.01  & 11.49  & 0.04  & 12.15  & 0.33  \\
J20472428$+$0018030 & E & 15.82  & 0.02  & 13.09  & 0.01  & 12.62  & 0.01  & 11.88  & 0.04  & 8.22  & 0.30  \\
\hline
\end{tabular}
\tablecomments{\hsize}{The columns are: (1) galaxy name; (2) galaxy type, ``S'' for Spiral and ``E'' for Elliptical; (3), (5), (7), (9) 7\;kpc aperture photometry result of $u$,$r$,$i$ and K$_s$-band, respectively, and (4), (6), (8), (10) their errors; (11) stellar age generated from $u$-$r$-$i$-K$_s$ color; (12) attenuation of V-band generated from $u$-$r$-$i$-K$_s$ color}
\end{table*}

\begin{table}
\caption[]{SFGs with strong enhancement\label{tab:3}}
\begin{tabular}{lrrcccrrrr}
\hline\hline
name & $z$ & $\rm sSFR_{enh}$ & pair\_type & bulge\_type & merge\_stage & $\rm \delta v$ & $\rm \log M_{HI+H2}$ & $\rm f_{8\mu m,nulear}$ & $\rm f_{8\mu m,total}$ \\ 
&  & (dex) &  &  &  & (km/s) & ($\rm M_{\odot}$) & (mJy) & (mJy)\\
(1) & (2) & (3) & (4) & (5) & (6) & (7) & (8) & (9) & (10)\\
\hline
J01183417$-$0013416 & 0.0453  & 1.34  & SS & p & early & 54 & 10.45  & 33.67  & 57.06 \\
J03381222$+$0110088 & 0.0392  & 0.75  & SE & c & early & 432.9 & 10.45  & — & — \\
J07543194$+$1648214 & 0.0459  & 0.71  & SS & p & late & 104.1 & 10.09  & — & — \\
J09155467$+$4419510 & 0.0396  & 1.11  & SS & p & late & 2.4 & — & — & — \\
J09155552$+$4419580 & 0.0396  & 1.36  & SS & p & late & 2.4 & — & — & — \\
J10100079$+$5440198 & 0.0460  & 0.79  & SS & c & late & 83.7 & — & 179.52  & 187 \\
J10332972$+$4404342 & 0.0523  & 0.84  & SS & p & early & 64.2 & — & 26.13  & 49.3 \\
J13153506$+$6207287 & 0.0306  & 1.59  & SS & p & early & 6 & 10.05  & 39.86  & 55.36 \\
J14005783$+$4251203 & 0.0327  & 0.82  & SS & p & early & 234.3 & 10.05  & 23.52  & 47.99 \\
J14005879$+$4250427 & 0.0335  & 0.95  & SS & p & early & 234.3 & 9.88  & — & — \\
J16024254$+$4111499 & 0.0335  & 0.77  & SS & c & early & 69 & 10.38  & — & — \\
J17045089$+$3448530 & 0.0572  & 0.80  & SS & c & late & 270 & — & 22.49  & 77.56 \\
J17045097$+$3449020 & 0.0563  & 1.04  & SS & p & late & 270 & — & 4.61  & 15.88 \\
\hline
\end{tabular}
\tablecomments{\hsize}{The columns are: (1) galaxy name; (2) redshift; (3) specific star formation rate enhancement; (4) pair type: ``SS'' for Spiral+Spiral pair, ``SE'' for Spiral+Elliptical pair; (5) bulge type: ``p'' for pseudo-bulge, ``c'' for classical-bulge; (6) interaction type \citep[detailed description in][]{2016ApJS..222...16C}; (7) difference in radial velocity between two galaxies in a pair; (8) HI gas mass in \citet{2018ApJS..237....2Z} plus H2 gas mass in \citet{2019A&A...627A.107L}, where the HI mass is divided by $\rm M_{star}$ ratio for SS pairs, and all assigned to the spiral for SE pairs; (9) Spitzer 8$\mu$m flux of the galaxy nuclear region \citep{2010ApJ...713..330X}; (10) Spitzer 8$\mu$m flux of the total galaxy}
\end{table}

\begin{table*}
\caption[]{Galaxies in late-stage S+E mergers\label{tab:4}}
\begin{tabular}{lrcrrrrrrr}
\hline\hline
name & $z$ & type & $\rm \log M_{star}$ & sSFR & $\rm \log M_{gas}$ & SDSS\_id & GROUP\_id & richness & rank \\ 
&  &  & ($\rm M_{\odot}$) &  & ($\rm M_{\odot}$) &  &  &  & \\
(1) & (2) & (3) & (4) & (5) & (6) & (7) & (8) & (9) & (10)\\
\hline
J10514368$+$5101195 & 0.0250  & E & 11.12  & -11.86  & 8.71 & 189845 & 1745 & 12 & 1 \\
J10514450$+$5101303 & 0.0238  & S & 10.85  & -11.64  & $<$8.66 & 189846 & 1745 & 12 & 2 \\
\\
J11542299$+$4932509 & 0.0702  & S & 10.98  & -11.35  & 9.49 & — & — & — & — \\
J11542307$+$4932456 & 0.0712  & E & 11.35  & -11.72  & $<$9.48 & 229360 & 215518 & 1 & 1 \\
\\
J13131429$+$3910360 & 0.0716  & E & 11.26  & -11.43  & $<$9.36 & 445348 & 47 & 103 & 1 \\
J13131470$+$3910382 & 0.0716  & S & 10.95  & -11.12  & $<$9.48 & 445349 & 47 & 103 & 2 \\
\\
J16354293$+$2630494 & 0.0701  & S & 11.27  & -11.31  & $<$9.52 & — & — & — & — \\
J16354366$+$2630505 & 0.0713  & E & 11.27  & -11.63  & $<$9.55 & 306304 & 177 & 49 & 1 \\
\hline
\end{tabular}
\tablecomments{\hsize}{The columns are: (1) galaxy name; (2) redshift; (3) galaxy morphology: ``S'' for Spiral, ``E'' for Elliptical; (4) stellar mass; (5) specific star formation rate; (6) gas mass from dust mass \citep{2016ApJS..222...16C}; (7) SDSS-ID; (8) Group-ID ; (9) richness (i.e. number of galaxies in the group); (10) rank (1=most massive, 2=other galaxies). Data in column (7)–(10) are from \citet{2012ApJ...752...41Y}. Those galaxies with no data in columns (7)–(10) are not included in the catalog of \citet{2012ApJ...752...41Y} for lack of spectroscopic redshift}
\end{table*}

\end{document}